\documentclass[fleqn,10pt]{wlscirep}
\usepackage[utf8]{inputenc}
\usepackage[T1]{fontenc}
\usepackage{appendix}
\usepackage{hyperref}
\usepackage{xr}
\usepackage{algorithm}
\usepackage{algpseudocode}   

\usepackage{lineno}

\usepackage[explicit]{titlesec}
\titleformat{\section}{\normalfont\Large\bfseries}{}{0em}{#1}
\titleformat{\subsection}{\normalfont\large\bfseries}{}{0em}{#1}
\titleformat{\subsubsection}{\normalfont\normalsize\bfseries}{}{0em}{#1}

\titlespacing\section{0pt}{12pt plus 4pt minus 2pt}{0pt plus 2pt minus 4pt}
\titlespacing\subsection{0pt}{12pt plus 4pt minus 2pt}{0pt plus 2pt minus 4pt}
\titlespacing\subsubsection{0pt}{12pt plus 4pt minus 2pt}{0pt plus 2pt minus 4pt}

\usepackage[capitalise,noabbrev,nameinlink]{cleveref}

\crefalias{figure}{figc}%
\crefname{figc}{Fig.}{Figs.}
\Crefname{figc}{Figure}{Figure}
\crefname{subfigure}{Fig.}{Figs.}
\Crefname{subfigure}{Figure}{Figures}

\Crefname{section}{Section}{Sections}
\crefname{section}{Section}{Sections}
\Crefname{subsection}{Subsection}{Subsections}
\crefname{subsection}{Subsection}{Subsections}

\newcommand{\secref}[1]{\textcolor{blue}{Section}  \textcolor{blue}{``\nameref{#1}''}}

\newcommand{\beginsupplement}{%
    \setcounter{table}{0}%
    \setcounter{figure}{0}%
    \setcounter{equation}{0}%
    \renewcommand{\theequation}{S\arabic{equation}}%
    \renewcommand{\thetable}{S\arabic{table}}%
    \renewcommand{\thefigure}{S\arabic{figure}}%
    \renewcommand{\tablename}{Supplementary Table}%
    \renewcommand{\figurename}{Supplementary Figure}%
    \crefalias{table}{apptab}%
    \crefalias{figure}{appfig}%
    \crefalias{section}{appsec}
    \crefalias{equation}{appeq}
}

\crefname{appfig}{Supplementary Fig.}{Supplementary Figs.}
\Crefname{appfig}{Supplementary Figure}{Supplementary Figure}
\crefname{apptab}{supplementary table}{supplementary tables}
\Crefname{apptab}{Supplementary Table}{Supplementary Tables}
\crefname{appsec}{supplementary section}{supplementary sections}
\Crefname{appsec}{Supplementary Section}{Supplementary Sections}
\crefname{appeq}{supplementary equation}{supplementary equations}
\Crefname{appeq}{Supplementary Equation}{Supplementary Equations}

\usepackage[makeroom]{cancel}
\usepackage{bbm}
\usepackage{amssymb,bm}  
\usepackage{enumitem}
\usepackage{graphicx}
\usepackage{subcaption}

\usepackage[acronyms,nonumberlist,nopostdot,nomain,nogroupskip,acronymlists={hidden}]{glossaries} 
\newglossary[algh]{hidden}{acrh}{acnh}{Hidden Acronyms}

\newcommand{\colorunderbrace}[3][red]{
    \colorlet{currentcolor}{.}
    {
        \color{#1}
        \underbrace{\color{currentcolor}{#2}}_{#3}
    }
}

\newcommand{\h}{H} 
\newcommand{\Sset}{\mathbb{S}} 
\newcommand{\Tset}{\mathbb{T}} 

\title{Temporal Point Process Modeling of Aggressive Behavior Onset in Psychiatric Inpatient Youths with Autism}

\author[1,*]{Michael Potter}
\author[1]{Michael Everett}
\author[1]{Ashutosh Singh}
\author[1]{Georgios Stratis}
\author[2]{Yuna Watanabe}
\author[1]{Ahmet Demirkaya}
\author[1]{Deniz Erdogmus}
\author[3,+]{Tales Imbiriba}
\author[2,+]{Matthew S. Goodwin}

\affil[1]{Northeastern University, Electrical and Computer Engineering, Boston, 02115, USA}
\affil[2]{Northeastern University, Bouvé College of Health Sciences and Khoury College of Computer Sciences, Boston, 02115, USA}
\affil[3]{University of Massachusetts - Boston, Computer Science, Boston, 02125, USA}

\affil[*]{potter.mi@northeastern.edu}
\affil[+]{these authors contributed equally to this work}


\begin{document}
\newacronym{qq}{QQ}{Quantile-Quantile}
\newacronym{asd}{ASD}{Autism Spectrum Disorder}
\newacronym{mcmc}{MCMC}{Monte-Carlo Markov-Chain}

\newacronym[
  plural=HPPs,
  longplural=Homogeneous Poisson Processes
]{hpp}{HPP}{Homogeneous Poisson Process}
\newacronym[
  plural=NHPPs,
  longplural=Non-Homogeneous Poisson Processes
]{nhpp}{NHPP}{Non-Homogeneous Poisson Process}
\newacronym[
  plural=HawkesPPs,
  longplural=Hawkes Point Processes
]{hwkpp}{HawkesPP}{Hawkes Point Process}
\newacronym[
  plural=HawkesExpPPs,
  longplural=Hawkes Point Processes with Exponential Kernel
]{hwkexp}{HawkesExpPP}{Hawkes Point Process with Exponential Kernel}
\newacronym[
  plural=Hawkes2ExpPPs,
  longplural=Hawkes Point Processes with Two Exponential Kernels
]{hwk2exp}{Hawkes2ExpPP}{Hawkes Point Process with Two Exponential Kernels}
\newacronym[
  plural=HawkesPLPPs,
  longplural=Hawkes Point Processes with Power Law Kernel
]{hwkpl}{HawkesPLPP}{Hawkes Point Process with Power Law Kernel}

\newacronym{psis-loo}{PSIS-LOO}{Pareto Smoothed Importance Sampling - Leave One Out}
\newacronym[
  plural=TPPs,
  longplural=Temporal Point Processes
]{psis}{PSIS}{Pareto Smoothed Importance Sampling}
\newacronym{elpd}{ELPD}{Expected Log Density}
\newacronym{hmc}{HMC}{Hamiltonian Monte Carlo}
\newacronym{cdf}{CDF}{Cumulative Density Function}
\newacronym{rtc}{RTC}{Random Time Change}
\newacronym{aic}{AIC}{Akaike Information Criterion}
\newacronym{bic}{BIC}{Bayesian Information Criterion}
\newacronym{nuts}{NUTS}{No U-Turn Sampler}
\newacronym{pdf}{PDF}{Probability Density Function}
\newacronym[
  plural=TPPs,
  longplural=Temporal Point Processes
]{tpp}{TPP}{Temporal Point Process}
\newacronym{iid}{IID}{independent and identically distributed}
\newacronym{is}{IS}{Importance Sampling}
\newacronym{ai}{AI}{Artificial Intelligence}
\newacronym{ml}{ML}{Machine Learning}
\newacronym{dl}{DL}{Deep Learning}
\newacronym{cnn}{CNN}{Convolutional Neural Network}
\newacronym{pl}{PL}{Power Law}
\newacronym{mh}{MH}{Metropolis Hastings}
\newacronym{ess}{ESS}{Effective Sample Size}
\newacronym{mcse}{MCSE}{Monte-Carlo Standard Error}
\newacronym{kde}{KDE}{Kernel Density Estimators}
\newacronym{mape}{MAPE}{Mean Absolute Percent Error}

\newacronym{sib}{SIB}{Self-Injurious Behavior}
\newacronym{ed}{ED}{Emotion Dysregulation}
\newacronym{ato}{ATO}{Aggression Towards Others}

\newacronym{mle}{MLE}{Maximum Likelihood Estimation}
\newacronym{eda}{EDA}{Electrodermal Activity}
\newacronym{lstm}{LSTM}{Long Short Term Memory}
\newacronym{lr}{LR}{Logistic Regression}
\newacronym{svm}{SVM}{Support Vector Machine}
\newacronym{nn}{NN}{Neural Network}
\newacronym{wd}{WD}{Wasserstein Distance}
\newacronym{mc}{MC}{Monte Carlo}
\newacronym{gof}{GOF}{Goodness-of-Fit}
\newacronym{loo}{LOO}{leave-one-out}
\newacronym{rocauc}{ROC-AUC}{Area Under the Receiver Operating Characteristic Curve}

\begin{abstract}
\par
\vspace{-2em}
Aggressive behavior, including aggression towards others and self-injury, occurs in up to 80\% of children and adolescents with autism, making it a leading cause of behavioral health referrals and a major driver of healthcare costs. Predicting when autistic youth will exhibit aggression can be challenging due to their communication difficulties. Many are minimally verbal or have poor emotional insight. Recent advances in Machine Learning and wearable biosensing demonstrate the ability to predict aggression within a limited future window (typically one to three minutes) in autistic individuals. However, existing works don't estimate aggression onset probability or the expected number of aggression onsets over longer periods, nor do they provide interpretable insights into onset dynamics. To address these limitations, we apply \glspl{tpp}, particularly self-exciting Hawkes processes, to model the timing of aggressive behavior onsets in psychiatric inpatient autistic youth. We benchmark several \gls{tpp} models by evaluating their goodness-of-fit and predictive metrics. Our results demonstrate that self-exciting \glspl{tpp} more accurately capture the irregular and clustered nature of aggression onsets, especially compared to traditional Poisson models. These incipient findings suggest that \glspl{tpp} can provide interpretable, probabilistic forecasts of aggression onset along a time continuum, supporting future clinical decision-making and preemptive intervention.
\end{abstract}

\flushbottom
\maketitle
%
%
\thispagestyle{empty}

\glsresetall
\vspace{-3em}
\section{Introduction}

Autism is one of the most prevalent childhood disorders, affecting approximately 1 in 36 children \cite{maenner2023prevalence}. A significant proportion (up to 80\%) of autistic children and adolescents exhibit aggressive behaviors, including \gls{sib}, tantrums, meltdowns, property destruction, and \gls{ato}. \cite{hattier2011occurrence,kanne2011aggression,matson2014assessing}. These behaviors are among the leading causes of referral to behavioral health services \cite{arnold2003parent} and contribute substantially to healthcare costs \cite{croen2006comparison}. Many autistic youth struggle with emotional regulation and self-reporting of their internal states \cite{mazefsky2013emotion}. Thirty to forty percent are minimally verbal, and others experience difficulties with emotional insight and self-awareness \cite{tager2017conducting}. As a result, aggressive behaviors appear unpredictable, occur at irregular, seemingly random times,  and thus create barriers to accessing community resources, therapy, education, and clinical services. Families caring for children with autism often face increased stress, social isolation, and financial strain due to concerns about unexpected aggressive behaviors in different environments \cite{davis2008parenting,hodgetts2013home}. In addition, these behaviors can negatively affect support professionals, leading to increased compensation for work-related injuries, higher absenteeism, and staff turnover \cite{kiely1998violence}. The cumulative effect of these challenges can demoralize caregivers and clinicians, disrupt patient care trajectories, and, in severe cases, require home-bound or residential placement, reducing quality of life while increasing costs.

\gls{tpp} models capture irregular stochastic event times in various domains. \gls{tpp} methods are  widely applied in seismology \cite{ogata1988statistical}, finance \cite{filimonov2015apparent}, and epidemiology \cite{gatrell1996spatial}. In seismology, \glspl{tpp} model earthquake occurrences by capturing the stochastic nature of event times and locations while accounting for aftershock sequences and spatial clustering to estimate the probability of future seismic activity. In finance, they capture the irregular timing of discrete price changes, trade executions, and market orders while accounting for serial dependence of events, offering insight into market stability and volatility. In epidemiology, \glspl{tpp} model the spread of disease by treating infections as stochastic events that trigger secondary infections, helping to estimate transmission rate, predict outbreaks, and evaluate interventions.  

Despite their broad applicability, \glspl{tpp} have not yet been used to model aggressive behavior onsets in inpatient youths with autism. These onsets exhibit irregular timing and temporal clustering, with bursts occurring within short intervals. Compared to point-estimate classifier methods, \glspl{tpp} offer several key advantages. They generate probabilistic forecasts by sampling entire trajectories, account for irregular inter-onset intervals, and support causal inference and knowledge discovery \cite{yan2019recent}. 
Applying \gls{tpp} to aggressive behavior onsets in inpatient youths with autism  allows us to address questions such as ``How many onsets will occur between 5-10 minutes of an observation session?'' and ``Do inpatient youths with autism exhibit self-excitation in onset patterns?'' \cite{shchur2021neural}.  

Several studies have explored automated aggressive behavior detection in youths with autism using time-series data. \gls{cnn}-based models analyzing video streams \cite{fadhel2023detecting} have linked the likelihood of aggressive behavior to hand movement speed, while hybrid \gls{cnn}-\gls{lstm} models leveraging physiological signals from wearable biosensors \cite{khullar2021meltdown} have been used to detect meltdowns and tantrums. While these methods improve detection in autistic youth, they do not improve safety for clinicians, caregivers, or patients. Predicting when a subsequent aggressive episode will occur before it happens could enable just-in-time preventive interventions to mitigate harm.

Recent advances in \gls{ml} and \gls{dl}  demonstrate promising capabilities predicting imminent aggressive behavior in inpatient youths with autism. Studies have employed \gls{lr}, \gls{svm}, and \glspl{nn} to forecast aggressive episodes one to three minutes into the future using three minutes of prior real-time peripheral physiological data from wearable biosensors \cite{imbiriba2023wearable,10.1145/3240925.3240980,imbiriba2020biosensor}. While the results are promising, these methods do not estimate the probability of onsets over longer future time windows and only utilize partial histories of physiological data and prior onsets during prediction. Forecasting larger future time windows requires modeling the physiological data autoregressively. Additionally, using only three minutes of prior physiological and onset information disregards potentially valuable historical data. Enabling long-term and adaptable forecasting of aggressive behaviors could improve resource allocation and intervention planning for clinicians and caregivers who support autistic youths.

This paper uses \glspl{tpp} to forecast the number of proximal aggressive behavior onsets over longer periods than currently explored in the literature. Unlike discrete-time representations commonly used in time series applications, \gls{tpp} modeling treats inter-onset times as random variables. This approach enables precise temporal modeling without requiring a fixed time window to aggregate onsets, which can introduce discretization errors \cite{xiao2017wasserstein}. Prior work has applied a discretized \gls{nhpp} model to analyze skin conductance responses alongside observed regulatory behaviors, identifying statistical differences in the physiological states of autistic youth  \cite{chaspari2014non}. The study examined self, co-, and combined regulatory behaviors to assess dyadic physiological regulation patterns. However, it focuses on statistical hypothesis testing relating to observed interpersonal regulatory behaviors rather than modeling or predicting future behaviors.  

The present work applies \glspl{tpp} to model and predict aggressive behavior onsets in inpatient youths with autism. We evaluate modeling assumptions using \gls{gof} statistics such as \gls{qq} plots, raw residual plots, and \gls{wd}. Model performance is assessed through predictive performance metrics such as \gls{psis-loo} \gls{elpd}, \gls{mape}, and \gls{rocauc}. Additionally, we interpret \gls{tpp} model parameters for knowledge discovery, focusing on the branching factor of the \gls{hwkpp} to understand the degree of self-excitation. This foundational study analyzes different \gls{tpp} strategies, including homogeneous, non-homogeneous, and self-exciting \glspl{tpp},  demonstrating their potential in addressing complex questions about aggressive behavior in inpatient youths with autism. 
\section{Methods}

\subsection{Data and Participants}
\label{subsec:datapartitipants}
\noindent \textit{Sites and study protocol}:
This study is a secondary analysis of data reported in \cite{imbiriba2023wearable} acquired from psychiatric inpatients serially enrolled at clinical inpatient sites (Bradley Hospital, Providence, Rhode Island; Cincinnati Children’s Hospital, Cincinnati, Ohio; Western Psychiatric Hospital, Pittsburgh, Pennsylvania; and Spring Harbor Hospital, Portland, Maine) participating in the Autism Inpatient Collection (AIC). This prognostic study was designed to estimate and validate a predictive model and is reported following the Transparent Reporting of a Multivariable Prediction Model for Individual Prognosis or Diagnosis (TRIPOD) reporting guidelines. The Institutional Review Board (IRB) at each participating site reviewed and approved the study, including the AIC protocols and aggressive behavior prediction protocols. The sites and their IRBs are: Bradley Hospital (Providence, RI), Cincinnati Children's Hospital (Cincinnati, OH), Spring Harbor Hospital (Portland, ME), and Western Psychiatric Hospital (Pittsburgh, PA). IRB approval of the AIC extended to this study with an amendment. Guardians of all study participants provided informed consent and were remunerated. The IRB approved this retrospective study in compliance with the Health Information Portability and Accountability Act. All methods were performed in accordance with relevant guidelines and regulations following the Declaration of Helsinki. 
\\ \\
\noindent \textit{Inclusion/Exclusion criteria}: In total, 86 inpatients were enrolled. Inclusion criteria included confirmation of autism via research-reliable administration of the Autism Diagnostic Observation Schedule-2 (ADOS-2) (see \cite{imbiriba2023wearable}) and parent-reported, staff-reported, or staff-observed physical aggression or self-injurious behavior. Exclusion criteria included not having a parent proficient in English or the individual with autism having prisoner status. Of the 86 participants enrolled in the study, 16 (18\%) were excluded from the analysis: 8 were unable to wear the physiological biosensor, and 8 were discharged before any behavioral observation could occur. The final analytic sample consisted of 70 participants.
\\ \\
\noindent \textit{Participant statistics}:  Participants ranged in age from 5 to 19 years (M = 11.85, SD = 3.5), and were predominantly male (88\%), white (90\%), and non-Hispanic (92\%)~\cite{imbiriba2023wearable}. Nearly half of the participants (46\%) were minimally verbal, as determined by ADOS-2 Module 1 or 2, and 57\% met criteria for intellectual disability based on Leiter-3 global IQ scores (M = 72.96, SD = 26.12). The length of inpatient stay varied from 8 to 201 days (M = 37.28, SD = 33.95).
\\ \\
\noindent \textit{Observational procedures and behavioral coding}: Research staff performed observational coding in the inpatient units. At the same time, inpatient study participants with autism wore the commercially available and regulatory-compliant E4 biosensor by Empatica, Inc. on their non-dominant wrist. In the present study, we focus solely on annotations of operationally defined (see  \cite{imbiriba2023wearable}) aggressive behavior (i.e., \gls{sib}, \gls{ed}, \gls{ato}) episode's start (onset) and stop (offset) times within observation periods using a custom mobile application.

A total of 429 behavioral observation sessions were conducted between March 2019 and March 2020, with a median of 5 sessions per participant (IQR = 6). These sessions totaled approximately 497 hours of observation time (Median = 4.4 hours, IQR = 4.9 hours per session).

Across these sessions, 6,665 aggressive behavior episodes were annotated:
\begin{itemize}
  \item \textbf{Self-injurious behavior (SIB)}: 3,983 episodes (60\%); Median = 2 per participant (IQR = 23); Median duration = 1.97 sec (IQR = 4.25 sec)
  \item \textbf{Emotional Dysregulation (ED)}: 2,063 episodes (31\%); Median = 8 per participant (IQR = 27); Median duration = 10.09 sec (IQR = 20.08 sec)
  \item \textbf{Aggression towards others (ATO)}: 619 episodes (9\%); Median = 1 per participant (IQR = 8); Median duration = 2.31 sec (IQR = 3.75 sec)
\end{itemize}

To assess inter-rater reliability, 20\% of the total dataset was randomly selected and independently double-coded by two trained research staff members at each inpatient site. Agreement was high across all categories: SIB ($\kappa$ = .93), ED ($\kappa$ = .95), and ATO ($\kappa$ = .86). 

Additional details on the dataset are available in~\cite{imbiriba2023wearable} and summarized in 
\textcolor{blue}{Supplementary Table S6}.


\subsection{Temporal Point Processes} 
\glspl{tpp} provide a principled statistical framework for modeling the generative process of events over continuous time. In this study, we leverage \glspl{tpp} to model the timing of aggressive behavior onsets in autistic youth. A \gls{tpp} is a stochastic process that represents the occurrence of discrete events (in our case, aggressive behavior onsets) over a fixed time window $[0, T]$ \cite{bae2023meta}. A session, or realization, of a \gls{tpp}, is an ordered sequence of onset times, denoted as $S_i = \{ t_1^{(i)}, t_2^{(i)}, \dots, t_{J_i}^{(i)} \}$, where $J_i$ is the number of onsets in session $i$, and $t_j^{(i)}$ represents the time elapsed since session $i$'s start to when the $j$-th onset occurred. Equivalently, a \gls{tpp} can be formulated as a counting process $N(t)$, representing the cumulative number of onsets up to time $t$ in a given observation session. Formally, we define this as:
\begin{equation}
    N(t) = \max \{ j : t_j \leq t , t_j \in S \},
\end{equation}
where $N(t)$ tracks the number of onsets within $[0, t]$. For a realization \(S=\{t_1,t_2,\dots,t_J\}\) of a \gls{tpp}, the history up to (but not including) time \(t\) is $H_t = \{\, t_j \in S \mid t_j < t \,\}$. When an onset at time \(t_j\) is included in the history we write \(H_{t_j^+} = \{t_1,\dots,t_j\}\), and when it is excluded we write \(H_{t_j^-} = \{t_1,\dots,t_{j-1}\}\).

A \gls{tpp} is completely characterized by the conditional intensity function \(\lambda(t \mid H_t)\): it defines the cumulative intensity, the \gls{pdf} and \gls{cdf} of the next onset time, and the likelihood function of observed onsets. The conditional intensity function gives the instantaneous expected rate of onsets at time \(t\) given the past history. Formally, for $t>t_j$ in an interval following the last observed onset,
\begin{align}
    \lambda(t \mid H_{t_j^+})
    &= \lim_{\Delta t \to 0} \frac{\mathbb{E}\big[N(t+\Delta t)-N(t)\mid H_{t_j^+}\big]}{\Delta t}.
\end{align}
Equivalently, in terms of the conditional density \(f(\cdot\mid H_{t_j^+})\) and distribution \(F(\cdot\mid H_{t_j^+})\) of the next onset time,
\begin{align}
    \lambda(t \mid H_{t_j^+}) &= \frac{f(t \mid H_{t_j^+})}{1 - F(t \mid H_{t_j^+})} \qquad\text{for } t>t_j.
\end{align}

Intuitively, for an infinitesimal interval \(\Delta t\), \(\lambda(t\mid H_t)\,\Delta t\) is the approximate probability (or expected count) of an onset occurring in \([t,t+\Delta t)\) conditional on the observed history up to $t$. 

The conditional cumulative intensity function (\cref{eqn:condcumint}) is the  represents the expected number of onsets within the time window \( (t_j, t]\), conditioned on the history of onsets up to (and including) \(t_j\). The integrated conditional intensity function (\cref{eqn:cum}) is the average number of onsets that occur within the observation session $[0,t]$.

\noindent\begin{minipage}{.5\linewidth}
\begin{equation}
    \Lambda(t | H_{t_j^+}) = \int_{t_j}^{t} \lambda(s| H_s) ds \label{eqn:condcumint}
\end{equation}
\end{minipage}%
\begin{minipage}{.5\linewidth}
\begin{equation}
    \Lambda(t) = \int_{0}^{t} \lambda(s| H_s) ds \label{eqn:cum}
\end{equation}
\end{minipage}

\noindent The conditional \gls{pdf} (\cref{eqn:condpdf}) and conditional \gls{cdf} (\cref{eqn:condcdf}) of an onset occurring at and before time $t > t_j$, respectively, given the past history of onsets within the session, is described by the conditional intensity function:

\noindent\begin{minipage}{.5\linewidth}
\begin{equation}
    f(t|H_{t_j^+}) = \lambda(t|H_{t_j^+}) \exp \left( -\int_{t_j}^t \lambda(s|H_s) ds \right) \label{eqn:condpdf}
\end{equation}
\end{minipage}%
\begin{minipage}{.5\linewidth}
\begin{equation}
    F(t|H_{t_j^+}) = 1 - \exp \left( -\int_{t_j}^t \lambda(s|H_s) ds \right), \label{eqn:condcdf} \\
\end{equation}
\end{minipage}

The conditional \gls{pdf} and \gls{cdf} map the conditional intensity, often hard to interpret for decision making, into familiar probability quantities (a density and a cumulative probability), which are the standard inputs for decision‑making frameworks \cite{shachter1992decision}. To simplify notation, we use $*$ to denote the conditional intensity function, integrated conditional intensity function, conditional \gls{cdf}, and conditional \gls{pdf} conditioned on the history $H_t$, as $\lambda^*(t)=\lambda(t|H_t)$, $\Lambda^*(t) = \int_{0}^{t} \lambda^*(s) ds, f^*(t)=f(t|H_t)$, and $F^*(t)=F(t|H_t)$ respectively. 

Throughout this study we benchmark the following parametric \glspl{tpp}: the homogeneous Poisson process (\gls{hpp}), a non‑homogeneous Poisson process with a power‑law intensity (\gls{nhpp}), and several self‑exciting Hawkes variants (exponential, power‑law, and multi‑exponential kernels). Specifying a parametric conditional intensity yields closed‑form expressions for the conditional pdf, cdf, and integrated conditional intensity function, and concentrates estimation on a compact parameter vector $\theta$. Moreover, parametric \glspl{tpp} are data efficient unlike non-parametric \glspl{tpp}. We emphasize self‑exciting models because they naturally capture temporal clustering of onsets, and are consistent with clinical theories that rising distress and physiological arousal can trigger cascades of aggressive behavior in autistic youth \cite{lindsay2000antecedent,scarpa2015physiological,cohen2011assessing}. Throughout this manuscript, the conditional density $f(t \mid H_t, \theta)$, distribution $F(t \mid H_t, \theta)$, and conditional intensity $\lambda^{*}(t \mid \theta)$ are functions of the model parameters $\theta$. To lighten notation, the dependence on $\theta$ is omitted hereafter unless necessary.

\subsubsection{Data Preprocessing}
\label{subsec:datapreprocessing}
\begin{figure}[H]
    \centering
    \begin{subfigure}[b]{0.32\linewidth}
        \centering
        \includegraphics[width=\linewidth]{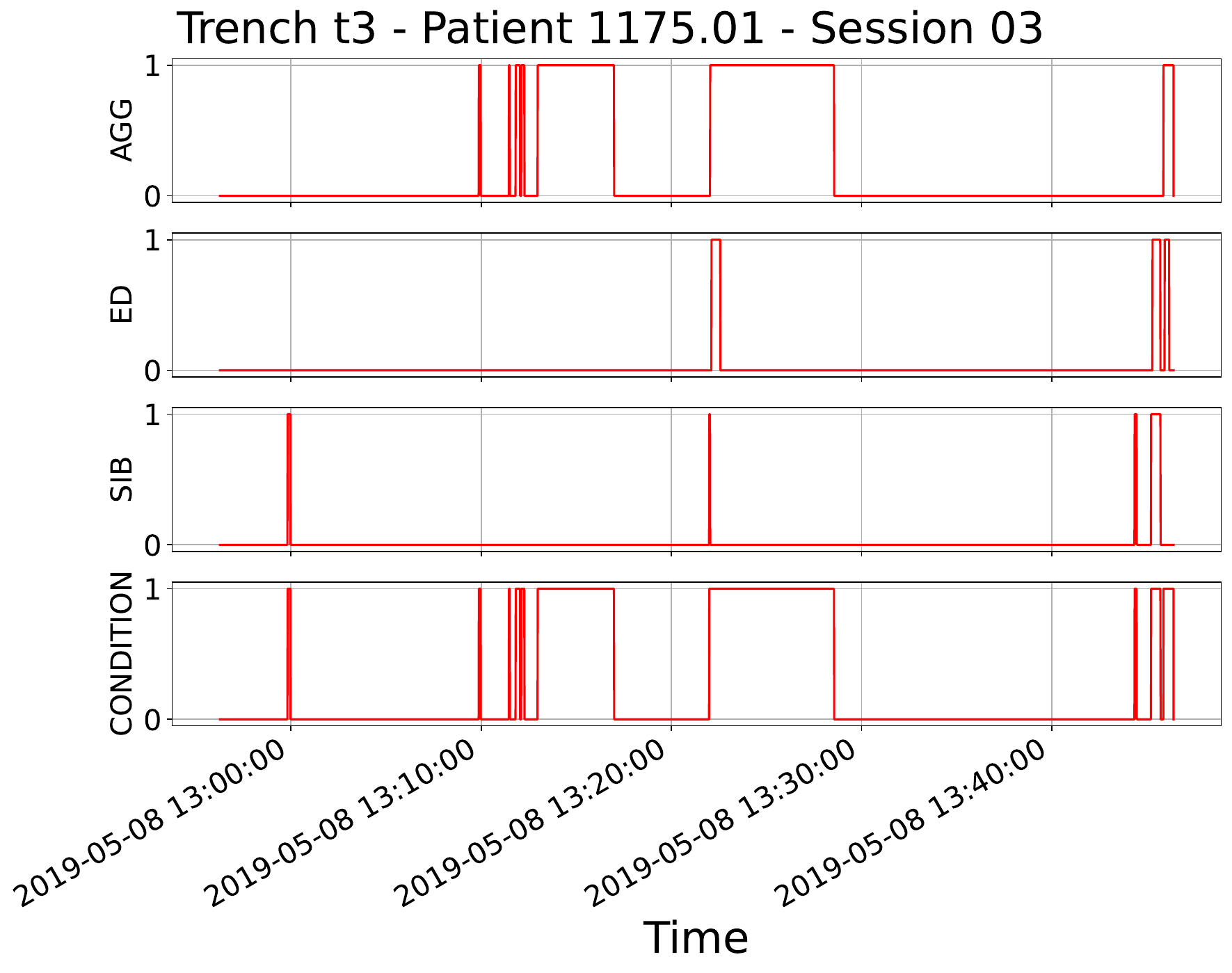}
        \caption{}
        \label{fig:dataaggregate}
    \end{subfigure}
    \hfill 
    \begin{subfigure}[b]{0.35\linewidth}
        \centering
        \includegraphics[width=\linewidth]{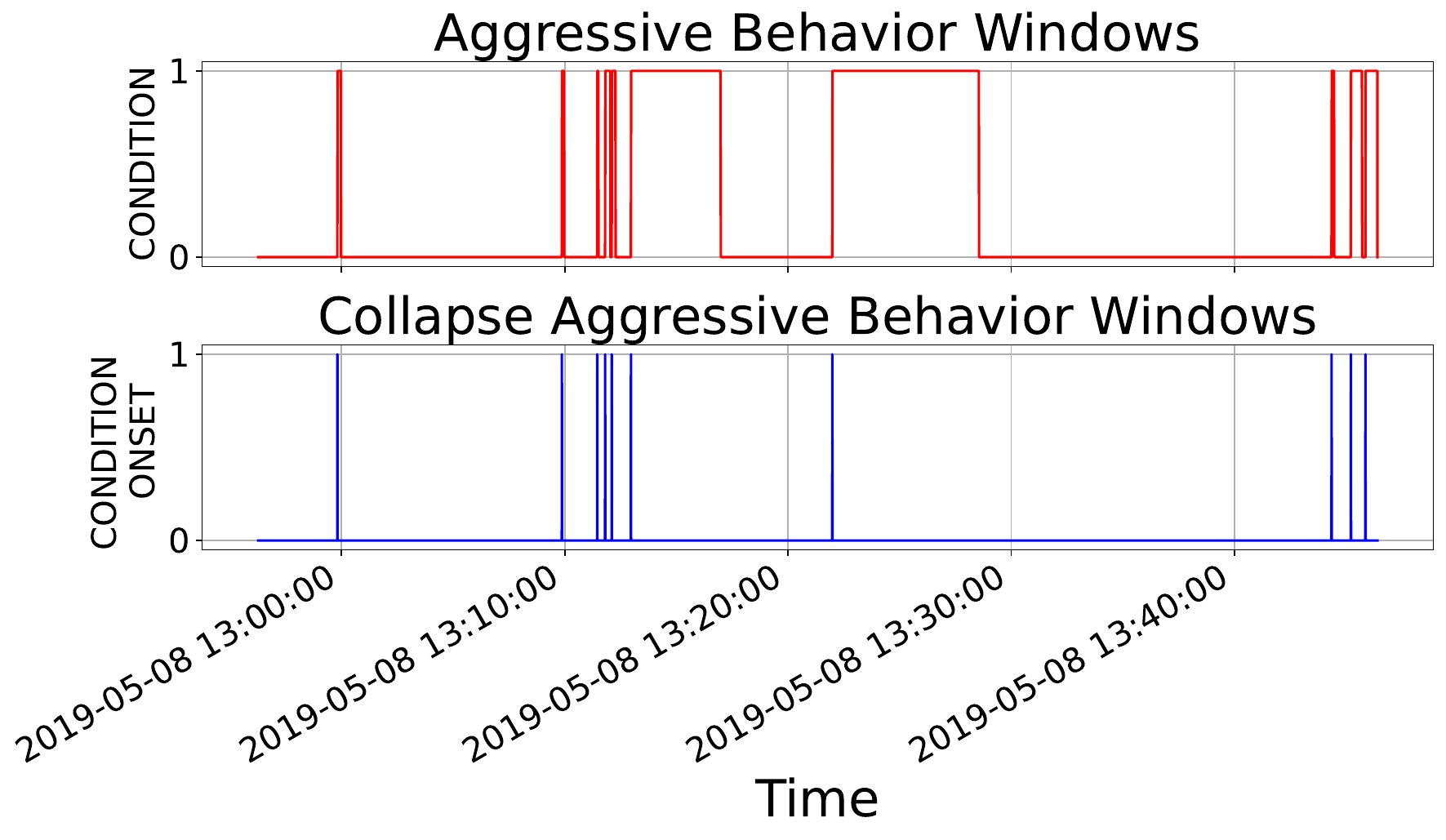}
        \caption{}
        \label{fig:datapreprocesscollapse}
    \end{subfigure}
    \hfill 
    \begin{subfigure}[b]{0.28\linewidth}
        \centering
        \includegraphics[width=\linewidth]{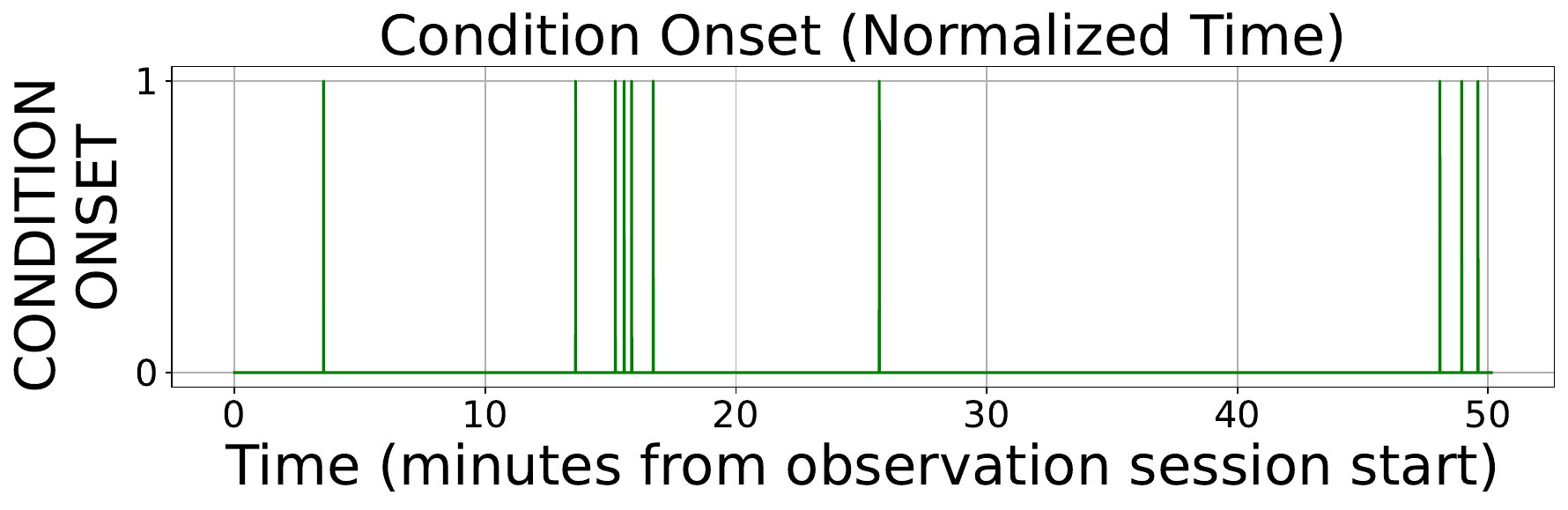}
        \caption{}
        \label{fig:datapreprocessnormalize}
    \end{subfigure}
    
    \caption{Three stages of data preprocessing. (a) Aggregating \gls{sib},\gls{ato},\gls{ed} behavior labels into a condition label. (b) Preprocessing only the onset timestamps. (c) Normalizing the onset timestamps to the beginning of the observation session.}
    \label{fig:preprocess}
\end{figure}

We treat \gls{sib}, \gls{ed}, and \gls{ato} as three recorded \glspl{tpp} but collapse them into a single event class for modeling (see \cref{fig:dataaggregate}). This is mathematically equivalent to the superposition of the behavior \glspl{tpp}. At the 250\,ms annotation resolution, we define an \emph{isolated event} as an episode consisting of a single positive annotation sample (i.e., episode length = 1 sample), whereas a \emph{continuous episode} is a run of consecutive positive samples. For analysis we reduce each continuous episode to its onset (the left boundary) and model only onset times and isolated events (that is, episodes of length one sample) (\cref{fig:datapreprocesscollapse}).

Onset timestamps were recorded in calendar time and are normalized per session by subtracting the session start time so each realization begins at 0 (\cref{fig:datapreprocessnormalize}). Observation sessions are treated as independent realizations; we do not concatenate sessions from the same participant because the unobserved gaps between sessions would bias parameter estimates and exacerbate edge effects \cite{filimonov2015apparent}.

\subsubsection{Likelihood Function}
The dataset comprises $I$ observed sessions $\Sset=\{S_i\}_{i=1}^I$ with corresponding observation lengths $\Tset=\{T_i\}_{i=1}^I$. For a given session $S_i$, the likelihood function is a joint density function of all onsets observed within the observation window $[0,T_i)$, factorized into conditional densities of each onset given all preceding onsets \cite{rasmussen2018lecture}:
\begin{align}
    L(\theta \mid S_i, T_i) = \big(1-F(T_i | H_{T_i}, \theta) \big) \prod_{j=1}^{J_i} f(t_j \mid H_{t_j}, \theta)
    \label{eqn:session_likelihood}
\end{align}
where the last term $(1-F(T_i|H_{T_i}))$ corresponds to the probability that no onsets are observed between $[t_{J_i},T_i)$. Using \cref{eqn:condcdf} and \cref{eqn:condpdf}, \cref{eqn:session_likelihood} becomes
\begin{align}
     L(\theta | S_i , T_i ) = \biggr( \prod_{j=1}^{J_i} \lambda^*(t_j \mid \theta) \biggr) \exp \left( -\int_0^{T_i} \lambda^*(s \mid \theta) ds \right) && t_0=0.
\end{align}
Then, the joint likelihood function across all observed sessions, assuming independence between sessions, is 
\begin{align}
    L(\theta \mid \Sset ,\Tset) = \prod_{i=1}^{I}  L(\theta | S_i , T_i ) \label{eqn:wholelikelihood}
\end{align}

In the following sections, we explore different choices for the conditional intensity function and their implications for onset analysis.

\subsection{Intensity Function Parameterizations}
\label{subsec:modelchoice}

Unlike typical \gls{ml} and \gls{dl} models, we employ several parametric \glspl{tpp} that are capable of modeling complex random sequences of onsets while remaining interpretable to clinicians. We aim to assess the feasibility of modeling and forecasting onsets using only the history of previous onsets within an observation session. Since our focus is on onset timing rather than additional attributes, we consider unmarked \glspl{tpp}, which rely solely on onset timestamps. For this study, we exclude physiological signals from our analysis, as work in progress using Shapley values \cite{nohara2022explanation} and stepwise regression \cite{zhang2016variable} suggests that the timing of aggression onset and offset is the key predictor of future aggression. This aligns with findings by Imbiriba et al\cite{imbiriba2023wearable} (see Fig.~1A-C in \cite{imbiriba2023wearable}), where \gls{rocauc} scores with physiological features alone are as low as 0.55, but exceed 0.80 when the most recent previous aggression onset is included as a covariate. Future work will incorporate physiological signals and contextual variables.

\subsubsection{Homogeneous Poisson Process}
The intensity function remains constant over time, meaning it does not change throughout the observation session. Additionally, the \gls{hpp} assumes independence from past onsets, implying that the timestamps of future onsets are independent of previous ones. Therefore, for a \gls{hpp}, the onsets within an observation session of length $T$ are distributed uniformly at random over the interval $[0, T]$, where the number of onsets in the session follows a Poisson distribution with rate $\mu$ \cite{gallager1997discrete}. The \gls{hpp} is described by the single parameter, $\mu$, giving the conditional intensity and integrated conditional intensity functions

\noindent\begin{minipage}{.5\linewidth}
\begin{equation}
    \lambda^*(t) = \mu
\end{equation}
\end{minipage}%
\begin{minipage}{.5\linewidth}
\begin{equation}
    \Lambda^*(t) = \mu t
\end{equation}
\end{minipage}
\newline

\noindent where $\theta= [ \mu ]$, $\theta \geq 0$, and  $\mu$ is the average number of onsets per unit time. 

\subsubsection{Non-Homogeneous Poisson Process}
The \gls{nhpp} has many parameterizations, but we choose the \gls{pl} characterization to describe conditional intensity and integrated conditional intensity functions. The conditional intensity function (\cref{eqn:nhpp_intensity}) is independent of past onsets, but it varies over time throughout the observation session. We choose the \gls{pl} parameterization because it is flexible enough to model a conditional intensity function which is either increasing, decreasing, or remaining constant over time, and has a closed-form integrated conditional intensity function (\cref{eqn:nhpp_icif}).

\noindent\begin{minipage}{.5\linewidth}
\begin{equation}
    \lambda^*(t) = \alpha k t^{k-1} \label{eqn:nhpp_intensity}
\end{equation}
\end{minipage}%
\begin{minipage}{.5\linewidth}
\begin{equation}
    \Lambda^*(t) = \alpha t^{k} \label{eqn:nhpp_icif}
\end{equation}
\end{minipage}

\noindent where    $\theta= [ \alpha, k ]$, $\theta \geq 0$, $\alpha$ is the scale parameter and $k$ is the shape parameter. The parameter $k$ controls the shape of the conditional intensity function: when $k > 1$, the intensity increases over time; when $k < 1$, the intensity decreases; and when $k = 1$, the intensity is constant, which reduces the model to the \gls{hpp}. The parameter $\alpha$ scales the overall conditional intensity function, affecting the steepness of the intensity decay or growth over time. 

\subsubsection{Self-Exciting Point Processes: Hawkes Process}
The \gls{hwkpp} is a self-exciting point process that removes the independence assumptions of traditional models mentioned previously by allowing future onsets to depend on past occurrences within an observation session. This dependence results in a  stochastic process, where both event times and intensity function are evolving random variables conditioned on the history of previous onsets \cite{laub2024hawkes}.

The conditional intensity function of a  \gls{hwkpp} is defined as:
\begin{equation}
    \lambda^*(t) = \colorunderbrace[blue]{\mu}{\text{baseline intensity}} + \colorunderbrace[red]{\sum_{t_j < t} \phi(t - t_j)}{\text{excitation trigger}}, \label{eqn:generalhwk}
\end{equation}
where $\mu$ represents baseline intensity, ensuring onsets occur even in the absence of prior onsets, and the summation term captures the self-exciting nature of the process. Another way to interpret this is that \(\mu\) models the exogenous onset occurrences, while the triggering kernel \(\phi(t)\) captures endogenous onsets. Each past onset $t_j$ contributes to the intensity at time $t$ through the triggering kernel $\phi(t - t_j)$, which is nonnegative and causal, meaning it only influences future onsets ($\phi(t) \geq 0$ for $t \geq 0$, and $\phi(t) = 0$ for $t < 0$). Typically, $\phi(\cdot)$ is chosen to be a monotonically decreasing function, ensuring the influence of past onsets diminishes over time.
This formulation models the cascading effect of aggressive behavior, where each onset increases the likelihood of subsequent onsets, with this effect gradually decaying. As a result, the \gls{hwkpp} captures the temporal clustering of onsets, making it a suitable framework for forecasting future occurrences based on historical patterns.

We evaluate two kernel functions of a \gls{hwkpp}: the Omori-Law (or \gls{pl}) kernel $\phi(t) = k (c+t)^{-p}$ and the exponential kernel $\phi(t) = \alpha \beta \exp(-\beta t)$. The \gls{hwkexp} is Markovian, enabling the conditional intensity at a given onset to be computed from the conditional intensity evaluated at the previous onset and the current onset. It is suitable for processes with rapid self-excitation decay, where the process quickly ``forgets'' past onsets. The parameter $\beta$ governs the delay density $\beta \exp(-\beta t)$, determining the time between excitations. A larger $\beta$ leads to faster decay of influence from previous onsets and shorter inter-event times, while a smaller $\beta$ results in slower decay and longer influence of past onsets \cite{hawkes1971point}. The parameter $\alpha$ is known as the infectivity factor, which is the average number of new onsets triggered by any given onset \cite{hawkes1971point}.

\noindent\begin{minipage}{.5\linewidth}
\begin{equation}
    \lambda^*(t) = \mu + \alpha \beta \sum_{t_j < t} \exp \left(-\beta (t - t_j) \right) \label{eqn:hwkci} 
\end{equation}
\end{minipage}%
\begin{minipage}{.5\linewidth}
\begin{equation}
    \Lambda^*(t) = \mu t + \alpha \sum_{t_j < t} \big(1 - \exp \left(-\beta (t - t_j) \right) \big) \label{eqn:hwkcci}
\end{equation}
\end{minipage}

\noindent where $\theta = [\mu,\alpha,\beta]$ and $ \theta \geq 0 $.

\noindent In contrast, the \gls{hwkpl} is used when self-excitation follows a heavy-tailed distribution, making it suitable for processes with long-term triggering effects, such as epidemics or earthquakes. Being non-Markovian, it requires more computation but exhibits scale-free behavior, effectively modeling long-range temporal dependencies and varying onset frequencies. The parameter $p$ controls the decay rate of self-excitation, with slower decay allowing past onsets to influence future occurrences over extended periods.  The time constant $c$ regularizes the behavior of the power law kernel at very short times \cite{filimonov2015apparent}.

\noindent\begin{minipage}{.5\linewidth}
\begin{equation}
    \lambda^*(t) = \mu + k \sum_{t_j < t} \left(c+t-t_j \right)^{-p} \label{eqn:hwkplci} 
\end{equation}
\end{minipage}%
\begin{minipage}{.5\linewidth}
\begin{equation}
    \Lambda^*(t) = \mu t + \sum_{t_j < t} \frac{k}{p-1} \left( c^{1-p} - (c+t-t_j)^{1-p}\right) \label{eqn:hwkplcci}
\end{equation}
\end{minipage}

\noindent where $\theta = [\mu,k,c,p]$ and $ [\mu,k,c] \geq 0  , p>1$.

The Omori-Law kernel can be approximated by a weighted sum of exponential kernels, combining the Markovian computation of the conditional intensity from the exponential kernel with the scale-free behavior of the Omori-Law kernel. We use a \gls{hwk2exp} model to capture multiple temporal patterns, such as high- and low-frequency onsets, which a single exponential kernel \gls{hwkpp} struggles to represent. By incorporating multiple kernels with different temporal scales, the multi-kernel \gls{hwkpp} models distinct behavior patterns, such as rapid bursts or sporadic aggression, improving onset prediction across timescales \cite{lee2024multi}. Adding more than two exponential kernels yields only marginal gains while increasing model complexity.

\noindent\begin{minipage}{.5\linewidth}
\begin{equation}
    \lambda^*(t) = \mu + \sum_{k=0}^1 \alpha_k \beta_k \sum_{t_j < t} \exp \left(-\beta_k (t - t_j) \right)
\end{equation}
\end{minipage}%
\begin{minipage}{.5\linewidth}
\begin{equation}
    \Lambda^*(t) = \mu t + \sum_{k=0}^1 \alpha_k \sum_{t_j < t} \big(1 - \exp \left(-\beta_k (t - t_j) \right) \big) 
\end{equation}
\end{minipage}

\noindent where $\theta=[\mu,\alpha_0,\beta_0,\alpha_1,\beta_1] $ and $\theta \geq 0$.

\begin{figure}[H]
    \centering
    \begin{minipage}{0.8\textwidth}
        \centering
        \includegraphics[width=\linewidth]{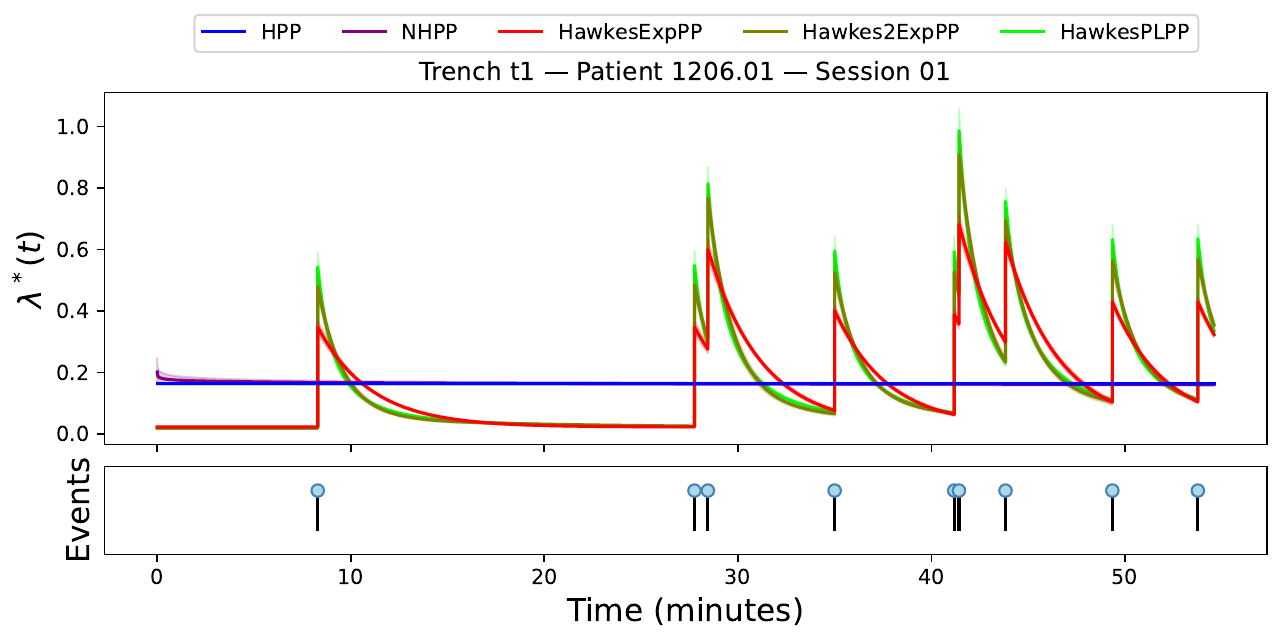}
        \subcaption{}
    \end{minipage}
    \begin{minipage}{0.8\textwidth}
        \centering
        \includegraphics[width=\linewidth]{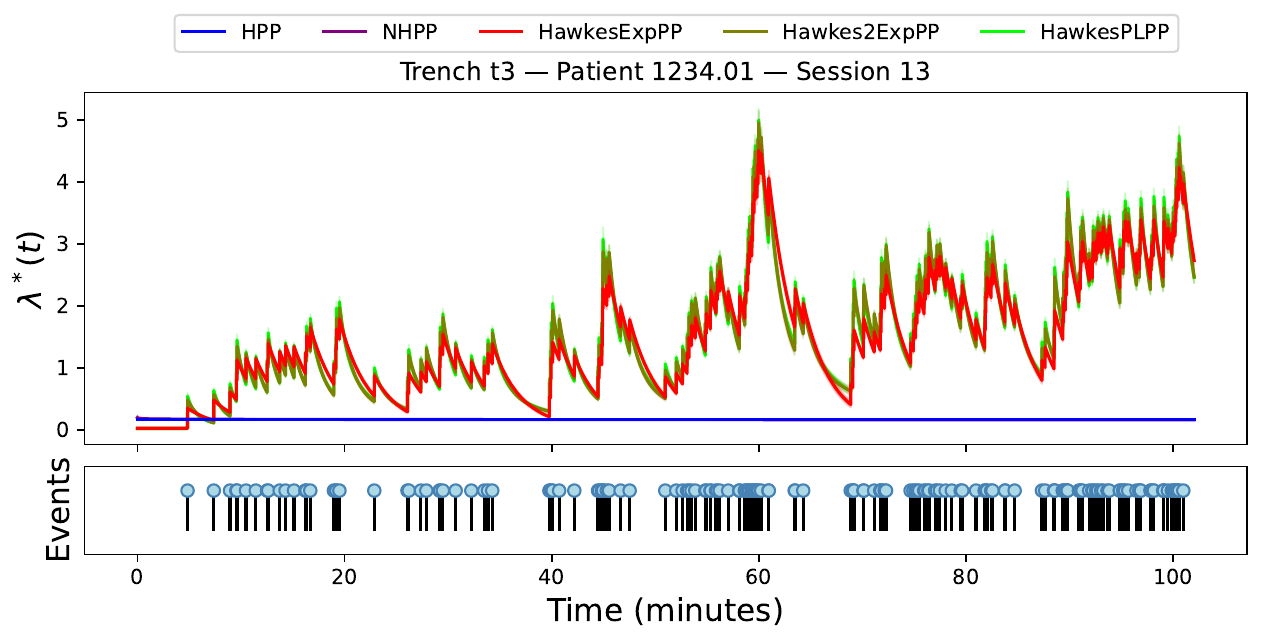}
        \subcaption{}
    \end{minipage}
    \begin{minipage}{0.8\textwidth}
        \centering
        \includegraphics[width=\linewidth]{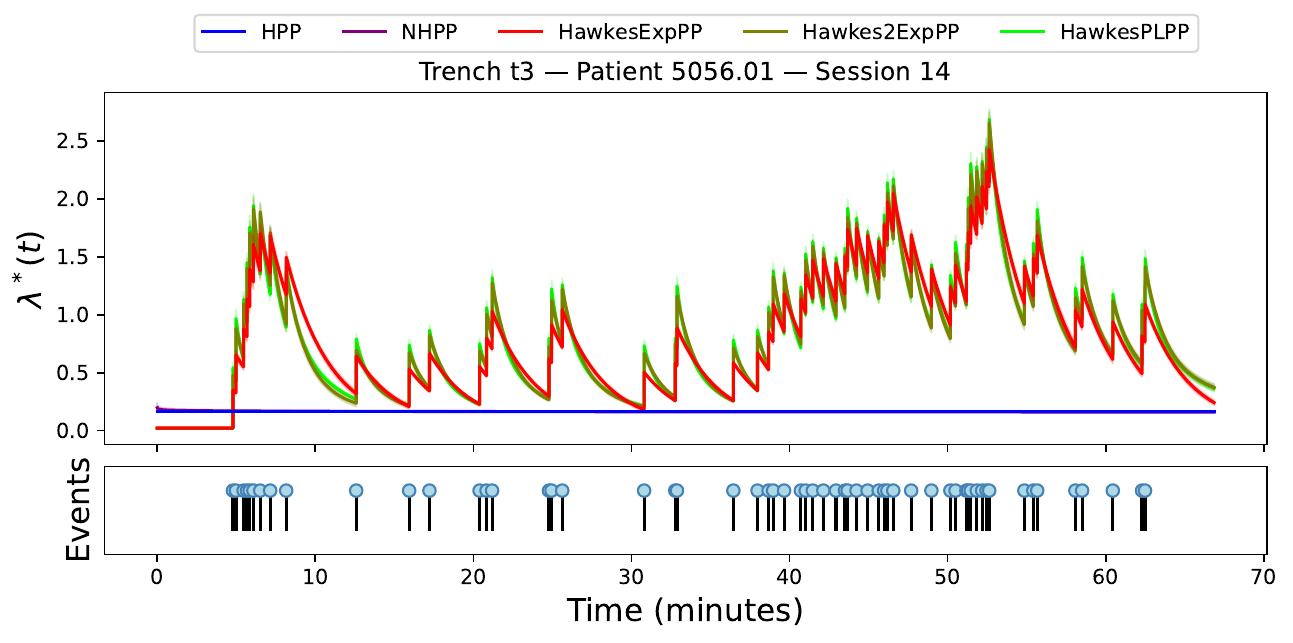}
        \subcaption{}
    \end{minipage}
    \caption{The median and 90\% credible interval of the \glspl{tpp} conditional intensity curves for two observation sessions.  The conditional intensity curve for each \gls{tpp} model in (a) participant 1206.01 - observation session 1, (b) participant 1234.01 - observation session 13, and (c) participant 5056.01 - observation session 14.}
    \label{fig:intensty_fns}
\end{figure}

\Cref{fig:intensty_fns} visualizes these intensity functions on our dataset to highlight their key differences.

We acknowledge that biases inherent to the data collection process, such as ``edge effects'' from finite observation session durations, are unavoidable and may impact model parameter inference. Edge effects in \glspl{tpp} arise when events before the observation window remain unobserved, leading to incomplete event histories. In \glspl{hwkpp}, this can result in underestimating self-excitation parameters and overestimating baseline intensity, as future events depend on past occurrences \cite{filimonov2015apparent}. However, if past events have only short-term influence, in keeping with our hypothesis that onsets influence proximal events, edge effects have minimal impact on parameter inference because the triggering kernel decays rapidly, making transient periods short. In preliminary analyses, we performed parameter inference using a modified exponentially decaying \gls{hwkpp} \cite{laub2024hawkes}:  
\begin{align}
    \lambda^*(t) = \mu + (\mu_0 - \mu) \exp(-\beta t) + \alpha \beta \sum_{t_j < t} \exp(-\beta (t-t_j) )
\end{align}
where the conditional intensity function starts at $\mu_0$. Unlike traditional formulations assuming a clear generative process start at \( t = 0 \), this approach acknowledges that many real-world processes have no well-defined beginning and are only observed from a certain point onward. Here, \( t=0 \) marks the start of the recording of the observation session rather than the process itself. Our preliminary analyses supported the assumption that edge effects does not meaningfully change the infectivity factor (no edge effect 0.896 versus edge effect 0.868) or time scale (no edge effect 0.362 versus edge effect 0.369); however, the baseline intensity $\mu$ decreases from $0.0229$ to $0.0168$. This outcome is expected, as edge effects, such as left truncation, are known to upwardly bias baseline intensity. Introducing $\mu_0$ for each session may help compensate for this bias. However, we find that the additional parameters provides only marginal benefits while introducing $I$ new parameters to the model (one $\mu_0$ per session). For this reason, we opted not to perform further experiments with this formulation and leave it to future work.

\subsection{Model Inference}
After preprocessing the data (\secref{subsec:datapreprocessing}), we perform Bayesian inference using \gls{mcmc} to estimate the posterior distribution of the \gls{tpp} model parameters. By Bayes' theorem, the posterior distribution of the model parameters $\theta$ given the observed sessions $\Sset$ and session durations $\Tset$ is  $\pi(\theta \mid \Sset, \Tset) \propto L(\theta \mid \Sset , \Tset) p(\theta)$,
where $L(\theta \mid \Sset , \Tset)$ is the likelihood function (\cref{eqn:wholelikelihood}), and $p(\theta)$ is the prior \gls{pdf}. We assign Gamma priors to all \gls{tpp} model parameters, except for the \gls{hwkpl}, as the Gamma distribution serves as a conjugate prior for the Poisson Process and offers flexibility over the positive real line. For the \gls{hwkpl}, we use wide uniform priors. To underscore how we chose weakly informative and wide prior \glspl{pdf}, the prior \gls{pdf} versus the posterior \gls{pdf} for the model parameters is show in \textcolor{blue}{Supplementary Figures S7 to S10}. Since direct computation of the posterior is often intractable due to high-dimensional integration or non-analytical integrals required for the partition function, we approximate it using \gls{mcmc}. \gls{mcmc} constructs an infinite dimension Markov chain whose stationary distribution converges to the target posterior: $p(\theta' | \Sset ,\Tset) = \int_\theta p(\theta' | \theta) p(\theta | \Sset, \Tset) d\theta$. The transition function, $p(\theta' | \theta)$ of the Markov Chain is denoted as the transition kernel.  Thus, each sampled state of the Markov chain represents a specific realization of the \gls{tpp} model parameters sampled from the posterior \gls{pdf}.

To generate samples from this Markov chain, we employ the \gls{nuts} \cite{hoffman2014no}, an adaptive variant of \gls{hmc}, as the transition kernel. In \gls{mcmc}, the transition kernel defines the probability of moving from one state to another while ensuring ergodicity and convergence to the posterior. Unlike random-walk-based methods, \gls{nuts} leverages Hamiltonian dynamics and gradients of the log-posterior to explore the posterior efficiently, proposing new states that are both distant from the current state (reducing autocorrelation in the \gls{mcmc} samples) and have high posterior probability (exploring the typical set of the posterior distribution \cite{betancourt2017conceptual}). For more details on the \gls{mcmc} sampling process, we refer readers to \cite{barber2012bayesian,hoffman2014no}. \gls{nuts}'s adaptive approach improves convergence and sampling efficiency in \gls{mcmc}, especially for posteriors with strong correlations, poor geometry, and nonlinear dependencies \cite{phan2019composable}. However, efficient mixing of the sampling chains and accurate posterior distribution sampling must still be validated using \gls{mcmc} diagnostics. 

The  \gls{ess} measures the number of effectively independent samples. The bulk \gls{ess} captures exploration of the high-density region of the posterior, while the tail \gls{ess} assesses sampling efficiency at the 5\% and 95\% quantiles \cite{carpenter2017stan} of the posterior. The \gls{ess} directly impacts the \gls{mcse}, which quantifies the uncertainty in the expectation estimates of the \gls{tpp} model parameters concerning the posterior distribution due to finite sampling: $ \text{MCSE} \approx \frac{\sigma}{\sqrt{\text{\gls{ess}}}} $, where $\sigma$ represents the standard deviation of the posterior samples. Higher \gls{ess} reduces \gls{mcse}, leading to more accurate estimates of the posterior mean of the \gls{tpp} model parameters. Conversely, low \gls{ess} increases \gls{mcse}, indicating strong autocorrelation in the samples. The Gelman-Rubin statistic ($\hat{R}$) compares within-chain and between-chain variances of the \gls{mcmc} sample chains to assess whether the Markov Chain explored the entire posterior distribution and has reached a stationary distribution. Values close to 1 suggest well-mixed chains. We also monitor divergences, which indicate numerical instabilities when the sampler struggles to explore the posterior, typically due to regions of high curvature or poor adaptation. 

We use 6000 warm up samples and collect 4000 post-warmup samples per \gls{mcmc} chain, and set a target acceptance probability of 0.99 for the Metropolis-Hastings acceptance criterion and \gls{nuts} kernel tuning \cite{barber2012bayesian,hoffman2014no}. In all experiments, we observe $\hat{R} \approx 1$ for all \gls{tpp} model parameters, zero divergences, high bulk and tail \gls{ess} for all \gls{tpp} model parameters, and low \gls{mcse} for all \gls{tpp}. Full \gls{mcmc} sampling diagnostic results are in \textcolor{blue}{Supplementary ``Monte-Carlo Markov Chain Diagnostics''}, \textcolor{blue}{Supplementary Tables S1 to S5}, and \textcolor{blue}{Supplementary Figures S1 to S5}.
These \gls{mcmc} diagnostics ensure efficient sampling of the posterior distribution; however, separate model validation metrics are required to assess the \gls{gof} and predictive performance of the selected parametric models.



\subsection{Model Validation}
\label{sec:modelvalidation}

A \gls{tpp} model for aggressive behavior onsets should effectively handle irregular inter-onset intervals, flexibly captures complex temporal dependencies among onsets within an observation session, maintains interpretability, and provides probabilistic forecasting of future onsets. Evaluation of these qualities relies on both \gls{gof} and forecasting performance metrics. \gls{gof} tools, such as \gls{qq} plots, raw residual plots, and count distributions, assess how well the \gls{tpp} captures the data-generating process of onsets \cite{wu2021diagnostics}. Forecasting performance is evaluated using metrics like \gls{psis-loo} \gls{elpd}, \gls{mape}, and \gls{rocauc}, which quantify the model’s ability to provide accurate probabilistic predictions of future onsets \cite{ludke2023add,vehtari2017practical}.

\subsubsection{QQ Plots}

A \gls{qq} plot is a residual analysis tool for comparing the quantiles of two distributions, often used to assess whether a dataset follows a specific theoretical distribution, such as an exponential distribution. The plot visualizes the relationship between observed data and the expected distribution by plotting the quantiles of the observed data against those of the theoretical distribution. If the scatter plot of the observed and theoretical quantiles closely follows a 45-degree line, it indicates strong alignment, suggesting that the observed data closely matches the theoretical distribution. To apply the \gls{qq} plot to \gls{tpp} models that are not \gls{hpp} and may have complex theoretical inter-onset distributions, we utilize the \gls{rtc} theorem \cite{brown1988simple}. The \gls{rtc} theorem asserts that, for a \gls{tpp} characterized by a conditional intensity function \(\lambda^*(t)\) defined over the history of aggressive behavior onset times \(t_1,\dots,t_J\), the transformed times \(\tau(t_1),\dots,\tau(t_J)\), obtained using the mapping function
\vspace{-0.5em}
\begin{align} 
\tau(t) = \int_0^t \lambda^*(s) ds = \Lambda^*(t), \label{eqn:rtc} 
\end{align}
will follow a standard Poisson process. In this case, the \gls{rtc} theorem inter-onsets $\tau(t_{j+1}) - \tau(t_j)$ are \gls{iid} and exponentially distributed with mean = 1 (representing one onset per transformed time unit). Since the true intensity function $\lambda^*(t)$ and integrated conditional intensity function $\Lambda^*(t)$ are unknown, we follow standard practice and use an estimate $\Lambda^*(t) \approx \Lambda^*(t \mid \bar{\theta})$ , where $\bar{\theta} = \mathbb{E}_{p(\theta \mid \Sset, \Tset)} [\theta]$. Thus, the \gls{qq} plot compares the empirical distribution of inter-onsets in the observed data after applying the \gls{rtc} theorem mapping (\cref{eqn:rtc}), while the theoretical distribution is an exponential distribution with mean = 1.

\subsubsection{Raw Residual Plots}
The raw residual plot is based on the Doob-Meyer decomposition theorem \cite{andersen2012statistical}. For a counting process \(N(t)\) with conditional intensity \(\lambda^*(t)\), the raw residual process is defined as
\begin{align}
M(t) = N(t) - \Lambda(t) \label{eqn:rawresidual}
\end{align}
where \( \Lambda(t) = \int_0^t \lambda^*(s) \, ds \) is the integrated conditional intensity function.
By the theorem, \(M(t)\) is a martingale with mean zero at all times \cite{wu2021diagnostics}. Therefore, the raw residuals should be centered at zero. To assess model fit, for each session we evaluated the raw residual process at the end of the observation period, \(M(T)\), where \(T\) denotes the session's end time, and plotted \(M(T)\) against the total number of onsets observed in that session, \(N(T)\). If the \gls{tpp} model fits well, the raw residuals should be symmetrically distributed around zero, indicating that the model accurately captures the expected number of onsets. Systematic deviations from zero would suggest that the model is either overestimating or underestimating onset counts, indicating a poor fit to the data \cite{wu2021diagnostics}.

\subsubsection{Count Distribution}
A well‑fitting \gls{tpp} should reproduce the empirical distribution of onset counts per session. For session windows $[0,T_i)$ let $N(T_i)$ denote the observed count. The empirical (true) count distribution is the distribution of $\{N(T_i)\}_{i=1}^I$. The model count distribution is obtained by simulating onsets in each $[0,T_i)$ using Ogata’s thinning algorithm \cite{laub2024hawkes}.
\vspace{-0.5em}
\begin{algorithm}[h!]
\caption{Posterior predictive count generation}
\label{algo:count_sampler}
\begin{minipage}{0.75\textwidth}
\begin{algorithmic}
\For{$T_i \in \Tset$}
    \State $\theta^{(s)} \sim p(\theta \mid \Sset, \Tset)$
    \State $\{t_1, t_2, \ldots, t_{n_i}\} \gets \textsc{OgataSample}\left(\lambda^*(t \mid \theta^{(s)}), T_i \right)$ 
    \Comment{stop sampling once $t_j \geq T_i$}
    \State $n_i \gets |\{t_1, t_2, \ldots, t_{n_i}\}|$
\EndFor
\State \Return $\{n_i\}_{i=1}^{I}$
\end{algorithmic}
\end{minipage}
\end{algorithm}
\noindent where \( \theta \) are the model parameter samples from \gls{mcmc}, \( t_j \) are sampled onset times conditioned on past history \( H_{t_{j-1}^+} \), and \( n_i \) is the total number of onsets within the window \( T_i \). The pseudo-code for generating samples of the onset counts per session is shown in \cref{algo:count_sampler}.  We then compare the empirical and simulated count distributions (e.g., via \gls{kde}) and quantify their discrepancy with the \gls{wd} \cite{kolourioptimal,ludke2023add}.

Unlike benchmark metrics such as log-likelihood, \gls{aic}, and \gls{bic}, that provide only relative model performance comparisons, \gls{qq} plots, raw residual plots, and count distribution plots assess how well the chosen model approximates the true generative process of onsets.

\subsubsection{Pareto Smoothed Importance Sampling - Leave One Out Expected Log Predictive Density}
Informally, the \gls{loo} \gls{elpd} evaluates the ability of a \gls{tpp} model to predict onset times in a new session based on other sessions:

\noindent\begin{minipage}{.5\linewidth}
\begin{equation}
    \text{ELPD}_{\text{loo}} = \sum_{i=1}^{N} \log p(S_i,T_i \mid \Sset_{-i}, \Tset_{-i}) 
\end{equation}
\end{minipage}%
\begin{minipage}{.5\linewidth}
\begin{equation}
    p(S_i , T_i, \mid \Sset_{-i}, \Tset_{-i}) = \int p(S_i , T_i \mid \theta) p(\theta \mid \Sset_{-i}, \Tset_{-i}) d\theta
\end{equation}
\end{minipage}

\noindent where $\Sset_{-i} = \Sset \backslash S_i$ represents all observation sessions except observation session $i$, and $\Tset_{-i} = \Tset \backslash T_i$ represents all session durations except duration $i$. A higher \gls{loo} \gls{elpd} value indicates a better predictive model. The \gls{psis-loo} \gls{elpd} approximates the \gls{loo} expected log predictive density using Pareto‑smoothed \gls{is}, avoiding refitting the model for each held‑out session. Using \gls{is}, the \gls{elpd} may be defined as 
\begin{align}
    p(S_i , T_i | \Sset_{-i}, \Tset_{-i}) \approx \frac{1}{\sum_{m=1}^M p(S_i, T_i \mid \theta^{(m)})}
\end{align}
where $M$ is the number of post-warmup \gls{mcmc} samples from the posterior distribution $p(\theta \mid \Sset)$.  However, raw \gls{is} can be unstable because the importance ratios may have high or infinite variance. The \gls{psis} procedure stabilizes the estimate by fitting a generalized Pareto distribution to the upper tail of the \gls{is} replacing extreme weights with Pareto‑smoothed quantiles. This reduces \gls{is} weight variance and produces more reliable \gls{is} \gls{elpd} estimates, especially when the \gls{is} weights are heavy‑tailed. We visualize each session  \gls{psis-loo} \gls{elpd} estimate normalized by the number of onsets in the session, $\log (p(S_i | \Sset_{-i})) / |S_i|$,  in \textcolor{blue}{Supplementary Figure S6}. However, log-likelihood values may be challenging to interpret; thus, we also include the \gls{mape} and \gls{rocauc} calculation for better interpretability.

\subsubsection{Mean Absolute Percentage Error}
To evaluate forecasting performance, we assess the \gls{tpp} model's ability to predict the number of onsets within a time window $[t,t+\Delta t]$. A starting point $t_{\text{start}} \in [\Delta t, T-\Delta t]$ is uniformly at random selected, with onsets in $[0,t_{\text{start}}]$ forming the history $\h$, and those in $[t_{\text{start}}, t_{\text{start}} + \Delta t]$ constituting the forecast goal. A session is not valid for a given $\Delta t$ if $T < 2 \Delta t$, as this would give an invalid uniform distribution to sample $t_{\text{start}}$ from. The \gls{mape} is calculated as
\begin{align}
\text{MAPE}(N(t,t+\Delta t), \tilde{N}(t,t + \Delta t)) = \frac{ | N(t,t + \Delta t) - \tilde{N}(t,t+\Delta t) | }{\max(1,N(t,t+\Delta t))} \times 100
\end{align}
where $N(t,t+\Delta t)$ is the observed number of onsets in $[t,t+\Delta t)$, and $\tilde{N}(t,t+\Delta t)$ is the median predicted count in $[t,t+\Delta t)
$ over $M=250$ samples, conditioned on the history $H_t$. Note that the denominator of \gls{mape} has $\max(1,N(t,t+\Delta t))$ to ensure that 0 is not a divisor. The median prediction is
\vspace{-0.5em}
\begin{align}
\tilde{N}(t,t+\Delta t) = \text{med}\left( \tilde{N}(t,t+\Delta t)^{(m)} \mid m = 1:M \right)
\end{align}
where each $\tilde{N}^{(m)}$ is sampled using the \gls{mcmc} posterior samples and Ogata's thinning algorithm for each $m$ realization. For each observation session $S_i$, we sample 25 uniformly at random $t_{\text{start}}$ locations and their corresponding forecasting windows to compute the \gls{mape}. A lower \gls{mape} indicates a better predictive model for onset counts within a future time window.

\subsubsection{Receiver Operating Characteristic Area Under the Curve}
\label{subsubsec:rocauc}
To evaluate classification performance, we assess the \gls{tpp} model's ability to predict whether at least one onset occurs within a time window $[t_{\text{start}}, t_{\text{start}}+\Delta t]$. Following the same windowing procedure as for \gls{mape}, a starting point $t_{\text{start}} \in [\Delta t, T-\Delta t]$ is uniformly at random selected, with onsets in $[0,t_{\text{start}}]$ forming the history $\h_{t_{\text{start}}}$, and those in $[t_{\text{start}}, t_{\text{start}} + \Delta t]$ constituting the forecast goal. A session is not valid for a given $\Delta t$ if $T < 2\Delta t$, as this would give an invalid uniform distribution to sample $t_{\text{start}}$ from. The probability of at least one onset occurring in the window is computed in closed form from the conditional \gls{cdf} (\cref{eqn:condcdf}) as
\begin{align}
P[t_{\text{start}} < t_{\text{onset}} \leq t_{\text{start}} + \Delta t \mid \h_{t_{\text{start}}}] = 1 - \exp\!\left(-\left(\Lambda^*(t_{\text{start}} + \Delta t) - \Lambda^*(t_{\text{start}})\right)\right)
\end{align}
where $\Lambda^*(t)$ is the integrated conditional intensity function (\cref{eqn:cum}). The ground truth label $y(t_{\text{start}}, t_{\text{start}}+\Delta t) \in \{0,1\}$ is 1 if any onset occurs in $[t_{\text{start}}, t_{\text{start}}+\Delta t]$ and 0 otherwise (\cref{fig:example_windowing}). The predicted score is the median posterior probability
\vspace{-0.5em}
\begin{align}
\tilde{P}(t_{\text{start}}, t_{\text{start}}+\Delta t) = \text{med}\!\left(P[t_{\text{start}} < t_{\text{onset}} \leq t_{\text{start}} + \Delta t \mid \h_{t_{\text{start}}}, \theta^{(m)}] \mid m = 1:M\right)
\end{align}
where each $\theta^{(m)}$ is a post-warmup \gls{mcmc} sample, and $M=4000$. For each observation session $S_i$, we sample 25 uniformly at random $t_{\text{start}}$ locations and their corresponding windows, then compute the \gls{rocauc} \cite{narkhede2018understanding} pooled across all sampled windows. A higher \gls{rocauc} indicates better discrimination between windows containing and not containing onsets.

\vspace{-1em}
\section{Results}

\subsection{Count Distribution}

\Cref{fig:countdistribution} compares the generated count distributions of aggressive behavior onsets per session across \gls{hpp}, \gls{nhpp}, \gls{hwkexp}, \gls{hwk2exp}, and \gls{hwkpl} models against the empirical count distribution. As shown in \cref{fig:countdistribution}(a,b), the empirical distribution (orange) is highly right-skewed with strong zero-inflation and a heavy tail, while the \gls{hpp} and \gls{nhpp} distributions (blue) appear truncated Gaussian-like, centered around 15-16 counts per session. For example, even in a 100-minute session with no onsets, the \gls{hpp} model predicts an average of 16 occurrences, underscoring its inability to capture zero-inflation. In contrast, the self-exciting \gls{tpp} models (\cref{fig:countdistribution}(c,d,e)) closely match the empirical distribution, with the \gls{kde} curves of observed and generated counts nearly overlapping for the \gls{hwkexp}. Quantitatively, \cref{tab:metric} reports the Wasserstein distance (\gls{wd}) over 250 \gls{mc} trials: the self-exciting processes substantially outperform \gls{hpp} and \gls{nhpp}, with \gls{hwkexp} achieving the smallest \gls{wd} of 2.28, more than four times lower than \gls{hpp}'s 9.7.


\subsection{QQ Plots}

\Cref{fig:qqplot} displays the \gls{rtc}-transformed inter-onset quantiles (\cref{eqn:rtc}) against the $\exp(1)$  theoretical quantiles for the \gls{hpp}, \gls{nhpp}, \gls{hwkexp}, \gls{hwk2exp}, and \gls{hwkpl} models. As shown in \cref{fig:qqplot}(a,b), the \gls{hpp} and \gls{nhpp} fail to capture the bursty nature of aggressive behavior onsets, allocating excessive probability to larger onset counts. This results in overestimation during low-intensity periods and underestimation during self-excitation phases, exposing their inability to model self-excitatory dynamics. \Cref{fig:qqplot}(c) shows that the \gls{hwkexp} tends to generate slightly shorter inter-onset times than observed, with the deviation most pronounced in the tail, indicating that it assumes subsequent onsets occur sooner than they do during extended gaps. The \gls{hwkpl} better captures the tail due to the scale-free, long-tailed nature of the \gls{pl} kernel. As shown in \cref{fig:qqplot}(d,e), both the \gls{hwk2exp} and \gls{hwkpl} correct this tail deviation: the \gls{hwk2exp} achieves this through two scaled exponential kernels capturing short- and long-term dependencies, while the \gls{hwkpl} has long-tail effects built in. 

\subsection{Raw Residual Plots}

\begin{figure}[H]
    \centering
    \begin{subfigure}[b]{0.95\linewidth}
        \centering
        \includegraphics[width=\linewidth]{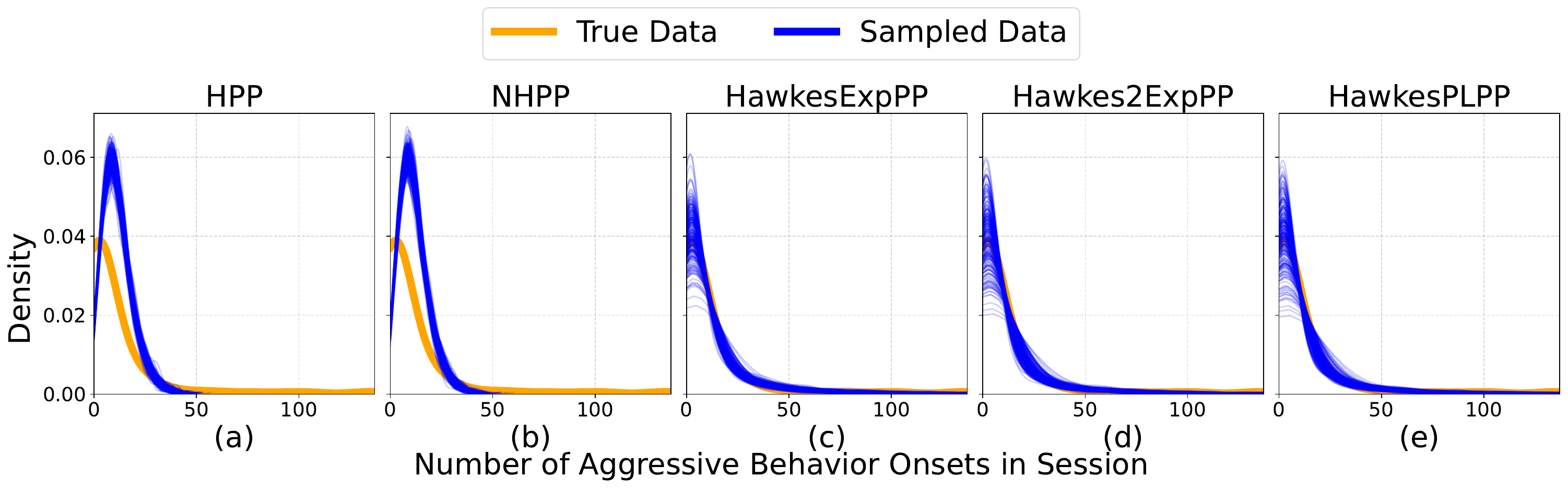}
        \caption{}
        \label{fig:countdistribution}
    \end{subfigure}
    
    \vspace{0.5cm} 
    
    \begin{subfigure}[b]{0.95\linewidth}
        \centering
        \includegraphics[width=\linewidth]{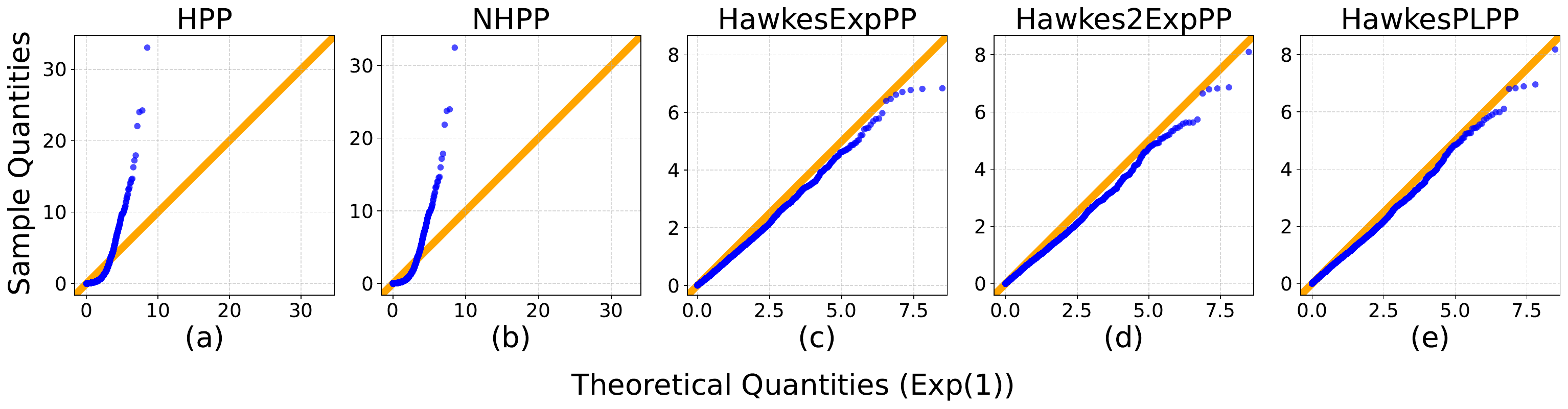}
        \caption{}
        \label{fig:qqplot}
    \end{subfigure}

    \vspace{0.5cm} 

    \begin{subfigure}[b]{0.958\linewidth}
        \centering
        \includegraphics[width=\linewidth]{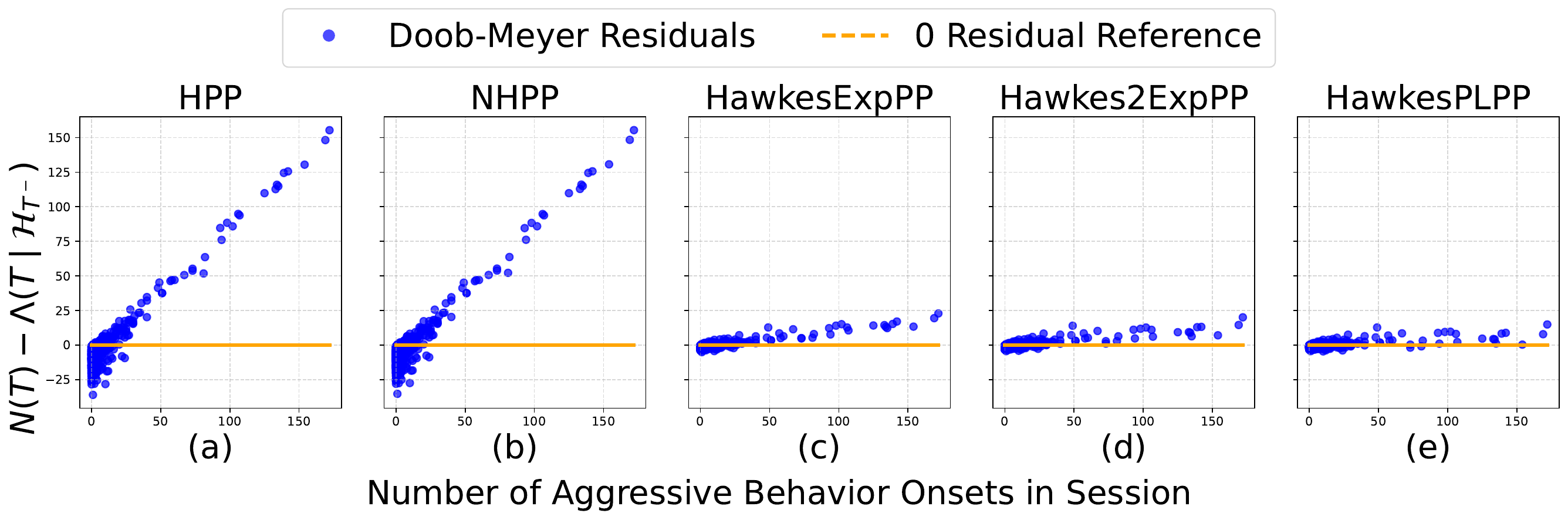}
        \caption{}
        \label{fig:rawresidual}
    \end{subfigure}

    \caption{Goodness-of-fit evaluation metrics for the \glspl{tpp}. Subfigure (a): The empirical data count distribution (orange) versus \gls{tpp} generated count distribution (blue) for the number of onsets in an observation session. Subfigure (b): the \gls{qq} plot of \gls{rtc} theorem inter-arrivals for different \gls{tpp} fits. The x-axis is the theoretical quantiles of an exponential distribution with mean 1, and the y-axis is the sample quantiles from all the \gls{rtc} theorem inter-arrivals in the dataset.  Subfigure (c): the scatter plot of the raw residuals (\cref{eqn:rawresidual}) versus the number of onsets in an observation session (blue). In each subfigure, (a)-(e) refer to the \gls{hpp}, \gls{nhpp}, \gls{hwkexp}, \gls{hwk2exp}, and \gls{hwkpl} models, respectively.}

    \label{fig:combined}
\end{figure}

\Cref{fig:rawresidual} plots the raw residuals (\cref{eqn:rawresidual}) against the number of onsets per session for each \gls{tpp} model. By the Doob-Meyer decomposition theorem, residuals,~\cref{eqn:rawresidual}, should be centered around 0 (orange line in \cref{fig:rawresidual}). \Cref{fig:rawresidual}(a,b) shows that the \gls{hpp} and \gls{nhpp} models overestimate onsets in sessions with few occurrences and underestimate them in bursty sequences, consistent with the \gls{qq} plots. In contrast, \cref{fig:rawresidual}(c,d,e) indicates that the \gls{hwkexp}, \gls{hwk2exp}, and \gls{hwkpl} residuals are closer to 0. However, they exhibit a slight positive drift in raw residual error as the number of onsets per session increases, indicating that the \glspl{hwkpp} tend to underestimate sessions with many onsets observed.


\subsection{Pareto Smoothed Importance Sampling - Leave One Out Expected Log Predictive Density}

For the \gls{psis-loo} \gls{elpd} estimates, the Pareto k diagnostic values indicated that \gls{psis-loo} provides a stable and accurate approximation to full \gls{loo} \gls{elpd} cross‑validation for our models. All Pareto k values fell in the ``good'' range for the self‑exciting models (\gls{hwkexp}, \gls{hwk2exp},\gls{hwkpl}) and for the fitted \gls{tpp} models overall. For the \gls{hpp} and \gls{nhpp} models, about 99\% of the Pareto k values were ``good'' and the remainder were ``ok,'' with a single \gls{nhpp} sample classified as ``bad.'' These categories follow the standard thresholds for the fitted Pareto k parameter used in \gls{psis-loo} diagnostics \cite{vehtari2017practical,vehtari2024pareto}. In terms of predictive performance, the \gls{hwkpl} model attained the highest \gls{psis-loo} \gls{elpd}, followed by the \gls{hwk2exp} and then the \gls{hwkexp}. The \gls{hpp} and \gls{nhpp} models showed substantially lower predictive performance: roughly a two‑fold reduction relative to the best self‑exciting models.

\subsection{Mean Absolute Percentage Error}

\noindent We evaluate the \gls{mape} for the forecasted number of onsets across time windows of $\Delta t = 1, 5, 10, 15, 20, 25$ minutes (\cref{tab:metric}). The performance gap between the \gls{hpp}/\gls{nhpp} and the self-exciting \glspl{tpp} widens as the forecasting window grows: the \gls{hwkexp} \gls{mape} rises from 11\% at 1 minute to 56\% at 25 minutes, while the \gls{hpp} exceeds 265\% at 25 minutes. Since the \gls{hpp} and \gls{nhpp} ignore onset history and self-excitation, they substantially overestimate future onsets in sessions with few or no onsets, whereas the \gls{hwkexp}, \gls{hwk2exp}, and \gls{hwkpl} models forecast fewer onsets under sparse history due to their low background intensity $\mu$.

Although \gls{mape} can be high for long-horizon probabilistic forecasts, \cref{fig:forecastcount} demonstrates the potential of estimating future onset counts from past onsets alone: the median forecast closely tracks the true counting process. The 90\% credible interval for the \gls{hwkexp} is notably asymmetric compared to the \gls{hpp}, likely reflecting the branching factor being near the critical regime ($\approx 1$) \cite{hardiman2013critical}, where values exceeding 1 yield nonzero probability of explosion. We discuss the branching factors of the \gls{hwk2exp} and \gls{hwkpl} in the later sections.

\subsection{Receiver Operating Characteristic Area Under the Curve}
\label{subsec:rocauc_results}

While \gls{mape} quantifies count-forecasting accuracy, a clinically meaningful complementary question is whether any aggressive episode will occur within a future window. \Cref{tab:rocauc} reports the \gls{rocauc} for predicting at least one onset within $\Delta t = 1, 5, 10, 15, 20, 25$ minute windows (see \secref{subsubsec:rocauc} for methodology, \cref{fig:example_windowing} for an example of positive and negative window labeling). The self-exciting models substantially outperform \gls{hpp} and \gls{nhpp} across all windows: the \gls{hwkpl} achieves \gls{rocauc} values of 0.85, 0.83, 0.81, 0.80, 0.78, and 0.78 across the six windows, while the \gls{hpp} hovers near 0.50, equivalent to random guessing, and the \gls{nhpp} performs only marginally better. This gap arises because \gls{hpp} and \gls{nhpp} ignore onset history when predicting future windows and cannot accommodate the temporal clustering observed in our data (\textcolor{blue}{Supplementary ``Ripley K Estimator''}). Notably, the \gls{hwkexp}, \gls{hwk2exp}, and \gls{hwkpl} achieve nearly identical \gls{rocauc} values, suggesting that the choice of triggering kernel matters less for binary classification than for count forecasting.

\subsection{Branching Factor}

The branching factor of a \gls{hwkpp} is the average number of ``child'' (endogenous) onsets triggered by each ``parent'' (exogenous) onset, defined as \( BF = \int_{0}^{\infty} \phi(t)\, dt \), where $\phi(t)$ is the triggering kernel in \cref{eqn:generalhwk}. The corresponding expressions for the \gls{hwkexp}, \gls{hwk2exp}, and \gls{hwkpl} are $\alpha$, $\alpha_0 + \alpha_1$, and $\frac{k}{p-1}c^{1-p}$ respectively. The branching factor governs process stability: $BF < 1$ is subcritical (stable), $BF = 1$ is critical, and $BF > 1$ is supercritical, yielding a potentially explosive cascade of onsets.

Estimating $BF$ requires care, as estimates for the \gls{hwkpl} and \gls{hwk2exp} can be inflated by outliers \cite{filimonov2015apparent}. Following \cite{filimonov2015apparent}, we compare quantile ratios (95\%/90\%, 99\%/95\%, and max/90\%) between synthetic and empirical inter-onset sequences using inverse \gls{cdf} interpolation for sessions with at least five onsets, defining outliers as inter-onset times exceeding $2\times$ the 99\% quantile. \Cref{tab:branchingfactor} shows no substantial evidence of outlier-induced bias, and the \gls{hwkexp} estimate is known to remain robust even under significant outlier contamination \cite{filimonov2015apparent}. Edge effects are another potential source of bias, but we assume their impact is minimal (see \secref{subsec:modelchoice}).

In the subcritical regime, $BF$ can be interpreted as the fraction of onsets triggered endogenously. The \gls{hwkexp} yields an average $BF$ of $0.85$-$0.95$, indicating that 85-95\% of onsets are triggered by past onsets rather than external factors. The \gls{hwk2exp} shows a small but nonzero probability of $BF$ slightly exceeding 1, and the \gls{hwkpl} shows a substantially larger one, indicating possible instances of exploding cascades of aggression within observation sessions.

\section{Discussion}
\label{sec:discussion}

Our results demonstrate that self-exciting point process models, such as the \gls{hwkpp}, fit the observed data well and capture the irregular and bursty nature of aggressive behavior onsets in psychiatric inpatient autistic youth. Compared to the standard \gls{hpp}, the Hawkes model substantially outperforms in both predictive accuracy and its ability to represent the underlying generative process.

When we frame the prediction task as a classification problem asking, ``Will any aggressive episode occur in the next 1, 5, 10, 15, 20, or 25 minutes?", the \gls{hwkpp} reliably distinguishes between windows where aggression is likely versus unlikely, while the standard \gls{hpp} cannot. We also note that our \gls{rocauc} values for predicting aggression within 1 to 5 minutes are very similar to those reported by Imbiriba et al.~\cite{imbiriba2023wearable} who included physiological features, further validating our modeling approach and results. Beyond classification, the \gls{hwkpp} model also provides more accurate forecasts of the actual number of aggressive episodes in each window. In practical terms, our model gives much more realistic estimates of how many aggressive episodes will occur in a given period, especially as the forecast window grows. Previous studies~\cite{imbiriba2023wearable,10.1145/3240925.3240980,imbiriba2020biosensor} cannot estimate the expected number of aggressive episodes in a given observation window. We demonstrated using \gls{wd}, a measure of how well a model captures the generative process of the real-world distribution of onset counts, that the \gls{hwkpp} model generates data that more closely matches the true pattern of aggressive behavior counts in clinical sessions compared to the traditional \gls{hpp}. These findings are further supported by the \gls{psis-loo} \gls{elpd}, which directly measures out-of-sample predictive performance. The Hawkes model’s \gls{psis-loo} \gls{elpd} is nearly twice as high (less negative) as the Poisson model’s, demonstrating that it generalizes substantially better to new sessions and is more reliable for forecasting onsets in real-world clinical settings.

Additionally, a key advantage of using \gls{tpp} models is their interpretability and explainability. For example, the inferred parameters for the Hawkes process show that after an initial onset, there is a high probability of subsequent onsets occurring in quick succession (a cascade effect). Thus, interventions that prevent the first instance or mitigate early occurrences, or ``mother'' onsets, should be prioritized. This suggests that careful interactions and interventions should be provided to inpatient youths with autism to prevent a cascade of subsequent onsets. These insights also introduce new research questions, such as whether we can generate the branching structure of the cluster representation of a \gls{hwkpp} to identify all the ``mother'' onsets and, through post-analysis, understand what triggers them.

\begin{figure}[H]
    \centering
    \begin{minipage}{0.49\textwidth}
        \centering
        \includegraphics[width=\linewidth]{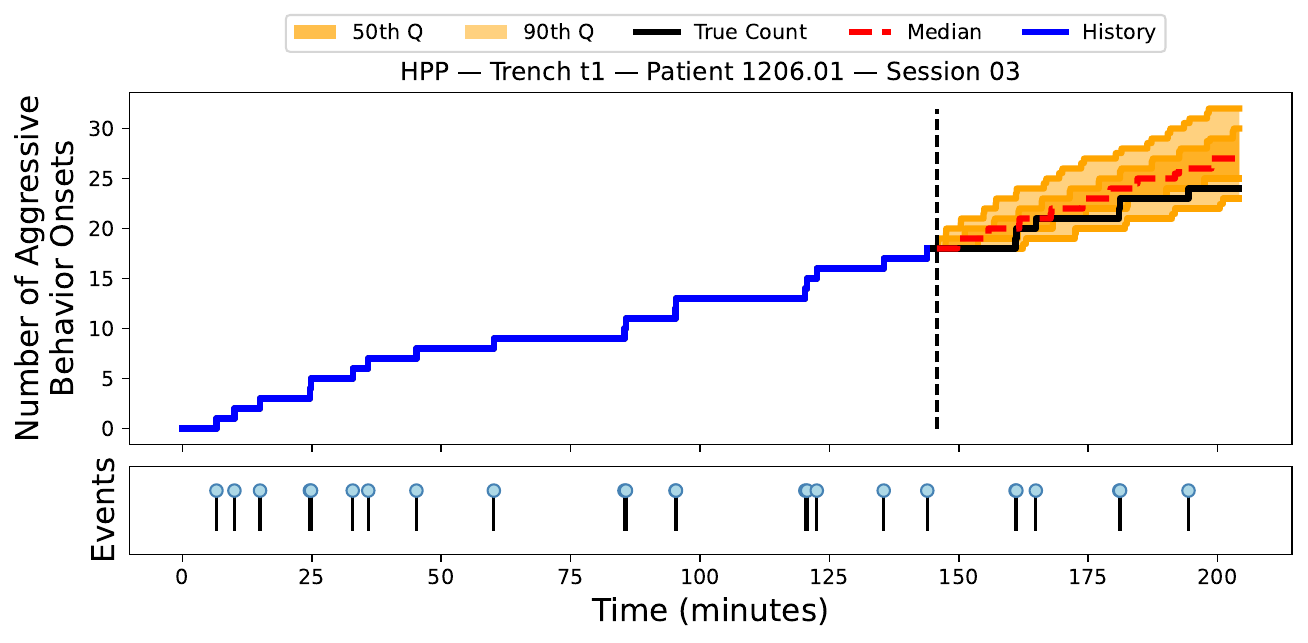}
        \subcaption{\gls{hpp}}
    \end{minipage}%
    \begin{minipage}{0.49\textwidth}
        \centering
        \includegraphics[width=\linewidth]{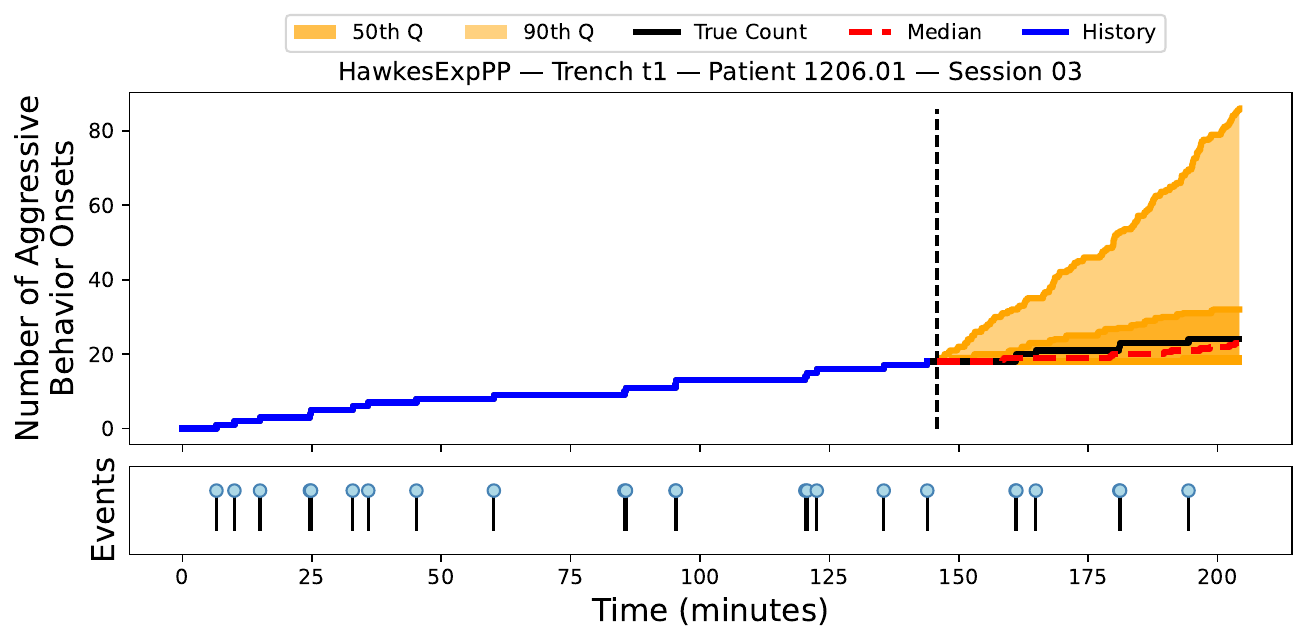}
        \subcaption{\gls{hwkexp}}
    \end{minipage}%
    
    \begin{minipage}{0.49\textwidth}
        \centering
        \includegraphics[width=\linewidth]{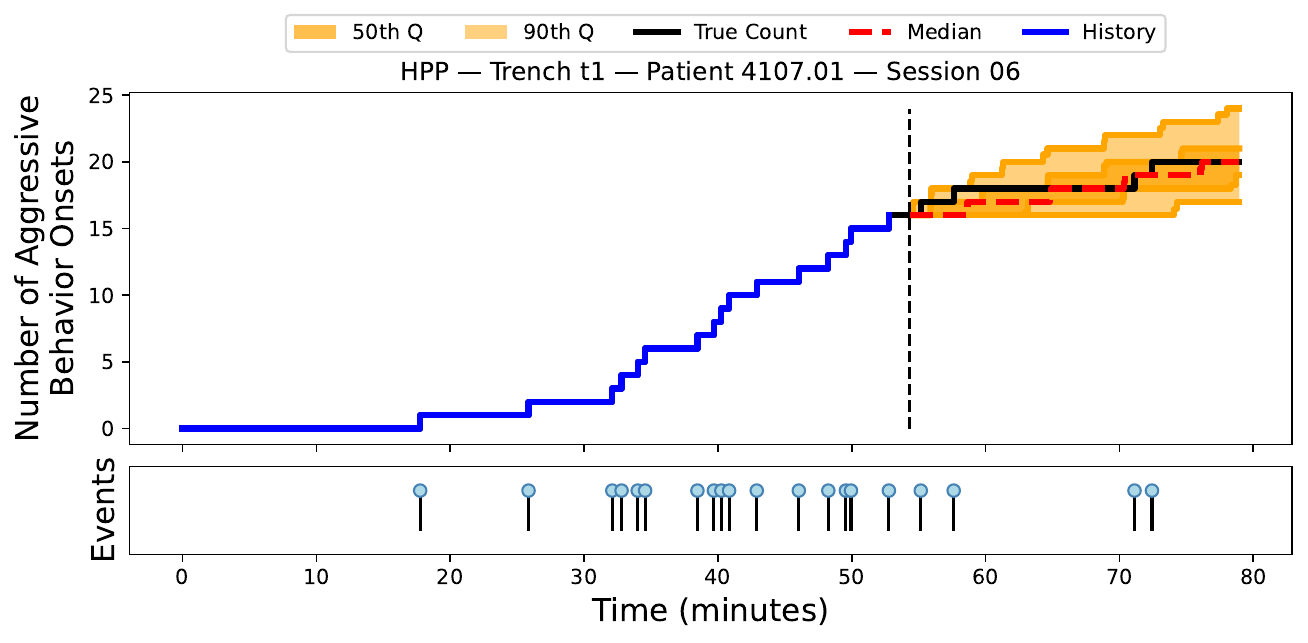}
        \subcaption{\gls{hpp}}
    \end{minipage}%
    \begin{minipage}{0.49\textwidth}
        \centering
        \includegraphics[width=\linewidth]{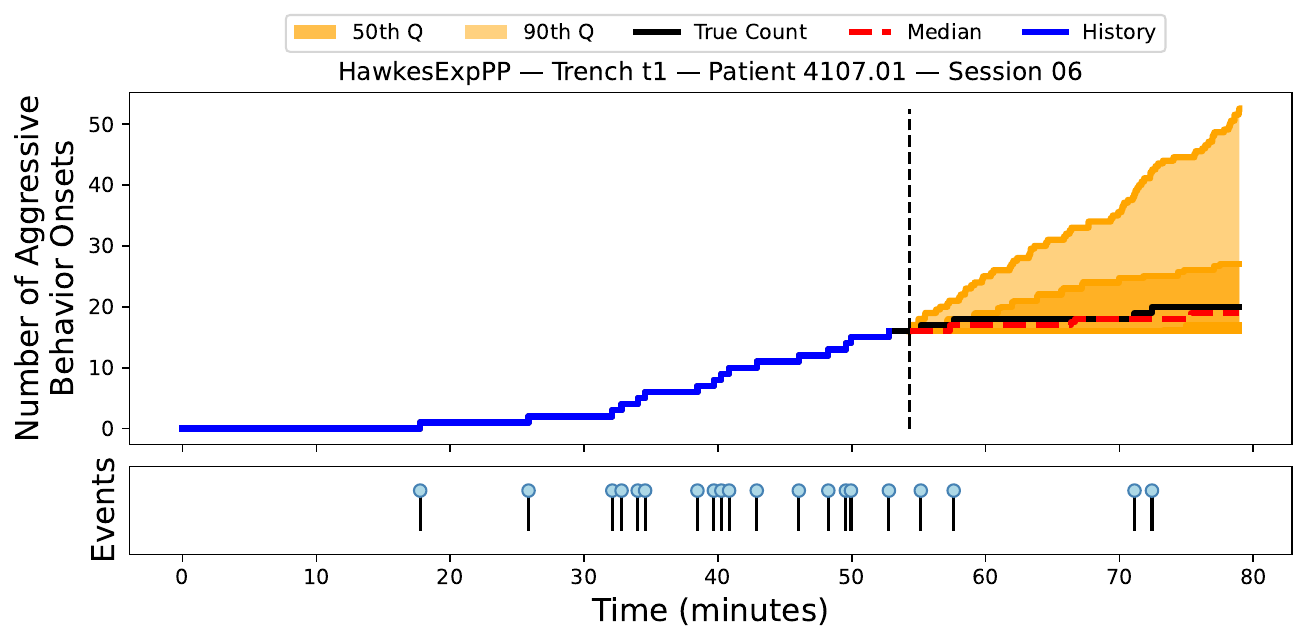}
        \subcaption{\gls{hwkexp}}
    \end{minipage}
    
    \begin{minipage}{0.49\textwidth}
        \centering
        \includegraphics[width=\linewidth]{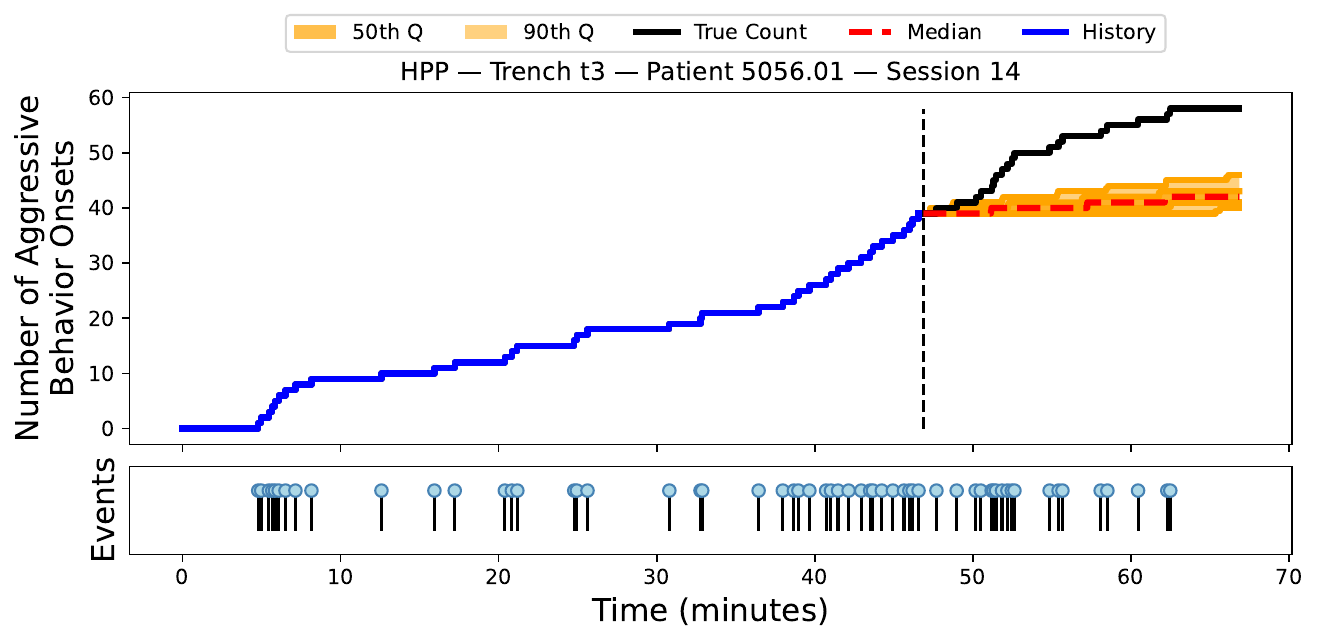}
        \subcaption{\gls{hpp}}
    \end{minipage}%
    \begin{minipage}{0.49\textwidth}
        \centering
        \includegraphics[width=\linewidth]{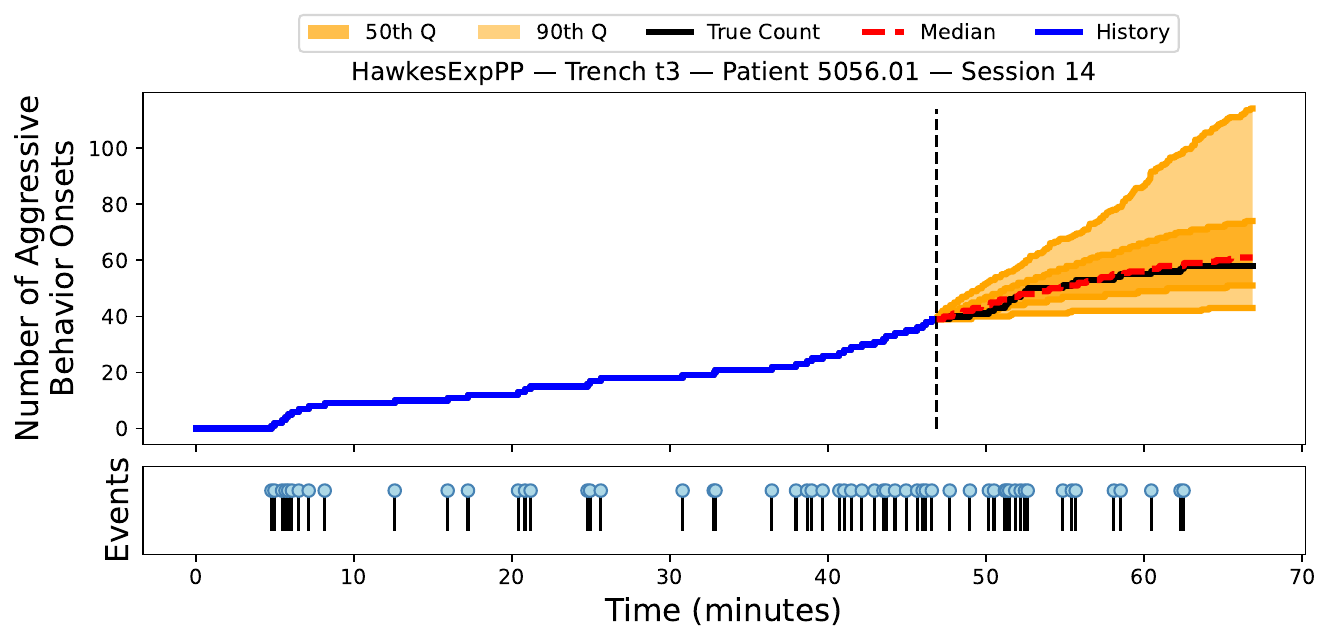}
        \subcaption{\gls{hwkexp}}
    \end{minipage}

    \caption{Visualization of the forecasted number of onsets within a future time window, conditioned on the history of prior onsets. The blue line represents the observed history of onset counts, while the black line denotes the true future onset count. The black vertical dotted line marks the start of forecasting. The red line corresponds to the median prediction over 250 sampled forecasts, with the darker orange shaded region indicating the interquartile range (25th to 75th quantile) and the lighter orange shaded region representing the 90\% credible interval (5th to 95th quantile). (a,b) show the forecasted number of onsets in participant 1206.01, session 3, (c,d) correspond to participant 4107.01, session 6, while (e,f) represent participant 5056.01, session 14, with each pair displaying forecasts for the \gls{hpp} and \gls{hwkexp}, respectively.  
}
    \label{fig:forecastcount}
\end{figure}

Our collective results demonstrate that our modeling approach improves predictive performance and provides potentially meaningful clinical insights and actionable forecasts. Accurately forecasting the number of onsets within a future window (\cref{fig:forecastcount}), predicting whether any onset will occur in that window (\cref{fig:probability_fns}), and estimating the precise timing of future onsets all have translational clinical potential. For example, the model’s predicted probability that at least one aggressive episode will occur in the next 5 or 10 minutes could be used to trigger real-time alerts for clinicians and caregivers, prompting increased supervision or preventive action. Forecasts of the expected number of onsets in a window could inform staff allocation and resource planning, ensuring that support is available when risk is highest. Precise timing predictions could allow for just-in-time adaptive interventions, such as de-escalation strategies or environmental modifications, to be delivered before an episode occurs. Additionally, the model’s ability to generate synthetic data that closely matches real-world patterns could enable safe simulation and testing of intervention protocols, as well as staff training, without risk to patients. Together, these outputs support proactive, data-driven decision-making to reduce harm and improve care in clinical settings.

Furthermore, our \gls{gof} visualizations which include the \gls{qq} plot (\cref{fig:qqplot}), raw residual plot (\cref{fig:rawresidual}), and count distribution (\cref{fig:countdistribution}) demonstrate that self-exciting point process models, such as the \gls{hwkpp}, can closely capture the complex, clustered onset patterns observed in psychiatric inpatient youth with autism. This strong fit is consistent with both expert clinical observation and prior analyses using the Ripley K estimator \cite{karlis2023proposer}, which highlight the temporal clustering of aggression in this population (\textcolor{blue}{Supplementary ``Ripley K Estimator''} and \textcolor{blue}{Supplementary Fig. S11}).
Predictive metrics, including \gls{psis-loo} \gls{elpd} and \gls{mape}, further confirm promise of \glspl{tpp} for forecasting aggression risk over clinically relevant time windows.

\section{Conclusion}
\label{sec:conclusion}
We applied \glspl{tpp}, particularly self-exciting Hawkes processes, to model the timing of aggressive behavior onsets in psychiatric inpatient autistic youth. To our knowledge, this is the first use of \glspl{tpp} in this population. Across 429 sessions and 6,665 annotated episodes from 70 participants, self-exciting models substantially outperformed homogeneous and non-homogeneous Poisson baselines on every metric. The \gls{hwkexp} reduced the Wasserstein distance between the empirical and model-generated count distributions to 2.28, more than four times lower than the \gls{hpp}'s 9.7. The self-exciting models also achieved \gls{psis-loo} \gls{elpd} values near $-7700$, nearly twice as high (less negative) as the \gls{hpp}'s $-13727$. Forecasting accuracy diverged sharply with horizon. At $\Delta t=25$ minutes, \gls{mape} was 55.96\% for the \gls{hwkexp} versus 269.52\% for the \gls{hpp}, a nearly five-fold reduction. For binary classification of whether any episode occurs in the next 1–25 minutes, the \gls{hwkpl} achieved \gls{rocauc} values of 0.85, 0.83, 0.81, 0.80, 0.78, and 0.78. The \gls{hpp} hovered near 0.50, statistically indistinguishable from random guessing.

Beyond predictive accuracy, the \gls{hwkpp} parameters yield clinically interpretable structure. The \gls{hwkexp} inferred a branching factor of $0.897 \pm 0.015$, with values ranging from $0.85$ to $0.95$. This indicates that 85–95\% of aggressive behavior onsets are endogenously triggered by prior onsets rather than exogenous factors, providing quantitative support for the clinical observation that early intervention on ``mother'' onsets can prevent cascades. Combined with closed-form probabilistic forecasts at clinically meaningful horizons, this interpretability distinguishes our approach from prior point-estimate classifiers. \glspl{tpp} are therefore a strong candidate framework for real-time clinical decision support, including risk-triggered alerts, staff-allocation forecasts, and just-in-time adaptive interventions to reduce harm and improve care for autistic youth.

\section{Limitations and Future Work}
Following this initial analysis, we intend to explore three future directions. First, we plan to divide our sample into sub-samples based on key variables such as the number of observation sessions, total observation time, and total onset count, duration, and intensity. By analyzing differences among these sub-samples, we aim to gain deeper insights into the variability of aggressive behavior and its recurrence. Second, we will investigate statistical differences in inferred model parameters and examine variations in modeling choices based on the type of condition label, including \gls{sib}, \gls{ed}, and \gls{ato}. Thirdly, we will utilize a hierarchical \gls{hwkpp} to account for between-person heterogeneity and disparities in data sample sizes among participants.

The distribution of onsets per participant appears to follow a power-law pattern. Specifically, one participant accounts for 26.5\% of the onsets (1,287 onsets), two participants contribute 25\% (475 to 910 onsets), five participants contribute 20.8\% (106 to 476 onsets), and 62 participants account for 27.7\% (0 to 105 onsets). This distribution suggests the existence of subpopulations within the data, conceptualized as chronically aggressive, hyper-aggressive, aggressive, and mildly aggressive. Given the substantial disparity in onsets exhibited by individual participants, hierarchical \gls{hwkpp} could address intra-individual and inter-individual variability. A single-level hierarchical \gls{hwkpp} could refine the analysis of the branching factor by modeling inter-individual heterogeneity and homogeneity. In contrast, a two-level hierarchical model would capture intra-individual and inter-individual session variability, further enhancing the analysis of edge effects, as discussed with the modified \gls{hwkpp} in \secref{subsec:modelchoice}. This two-level model would also enable the inference of a $\mu_0$ per observation session. Finally,  we aggregated three distinct behaviors into a single aggression onset label. A more nuanced analysis would involve modeling \gls{ato}, \gls{sib}, and \gls{ed} separately as each type of aggression may follow a distinct generative process, with potentially different baseline intensities and interacting infectivity factors. This modeling of interacting point processes could be achieved via a Multivariate Hawkes process \cite{liniger2009multivariate}.

While our approach demonstrates strong promise in learning a \gls{tpp} that closely captures the generative process of onsets, it still has limitations in predictive performance, mainly when answering questions like ``how many onsets will occur within the next 5 minutes?''. This modeling gap may be due to limited data sample size, the need for more expressive \gls{tpp} models (such as those with advanced kernel functions or deep learning-based architectures), or the exclusion of important covariates like physiological signals.

Furthermore, a limitation of generative modeling is that although we may learn the generative process of onsets with reasonable accuracy, this does not necessarily translate into improved predictive capability. To illustrate with a simple example, consider inferring the probability of heads in a fair coin toss. Estimating this probability as 0.5 from multiple flips is straightforward. However, even with an accurate estimate, it does not enable precise prediction of the outcome of the next flip or a sequence of flips. Our model may effectively capture the underlying generative process but may still struggle to predict proximal occurrence given the history of onsets within an observation session. Thus, while our models provide probabilistic forecasts that could support real-time clinical decision-making, there is still a lack of standardized, interpretable metrics for evaluating temporal point process performance in decision-based frameworks. In particular, metrics that are meaningful and actionable for clinicians \cite{shachter1992decision,shchur2021neural}. Addressing these challenges will be essential for translating \gls{tpp} modeling into practical, impactful tools for clinical care.

Lastly, a limitation towards real-time model parameter estimation is that \gls{mcmc} inference is notoriously slow, with online parameter update mechanisms still under research. However, we may explore online inference methods, such as Cubature Kalman Filters, for estimating the parameters of the intensity function in a \gls{hwkpp}. This would allow us to perform online updates of the model parameters with streaming data, bringing us a step closer to enabling just-in-time preventative interventions to reduce aggressive behavior in autistic youth.

\section{Software Packages}\label{sec:software_shoutout}

All analyses were conducted in Python, (version 3.9) following the \textit{Cookiecutter Data Science} project structure and workflow \cite{drivendata2025}. Bayesian inference was performed using \texttt{NumPyro=0.16.1} \cite{phan2019composable}, a probabilistic programming language built on \texttt{JAX=0.4.30} \cite{jax2018github} for efficient \gls{nuts} sampling. Model diagnostics (including \gls{psis-loo} cross-validation) and posterior exploratory analysis utilized \texttt{ArviZ=0.17.1} \cite{arviz_2019}, a toolkit for exploratory analysis of Bayesian models. Evaluation metrics such as \gls{rocauc} and \gls{wd} were calculated using \texttt{scikit-learn=1.5.2} and \texttt{scipy=1.12.0}, respectively. Data preprocessing employed \texttt{pandas=2.2.3}, \texttt{numpy=1.26.4}, and \texttt{xarray=2024.7.0}. Visualizations were created using \texttt{matplotlib=3.9.3}, and \texttt{seaborn=0.13.2}.

\pagebreak



\section{Acknowledgements}
\noindent This study was supported by grant R01LM014191 from the National Institutes of Health, the Simons Foundation for Autism Research Initiative, and the Nancy Lurie Marks Family Foundation. We are grateful to participating families. Families were offered \$75 for their child’s participation in the study. All project personnel listed were paid as research assistants for the duration of their involvement.

\section{Author Contributions Statement}
\noindent M.P.,A.S.,D.E.,T.I.,M.S.G conceived the experiments. M.S.G.,Y.W. conducted the data collection. M.P.,M.E.,T.I. conducted the experiments. M.P.,M.E.,T.I.,M.S.G. analyzed the results. All authors reviewed the manuscript. D.E., M.S.G obtained funding.


\section{Competing Interests}
\noindent The author(s) declare no competing interests.


\section{Data Availability}
\noindent The datasets generated during and/or analysed during the current study are available from the corresponding author on reasonable request.


\bibliography{sample}

\pagebreak

\begin{figure}[H]
    \centering
    \includegraphics[width=1.0\linewidth]{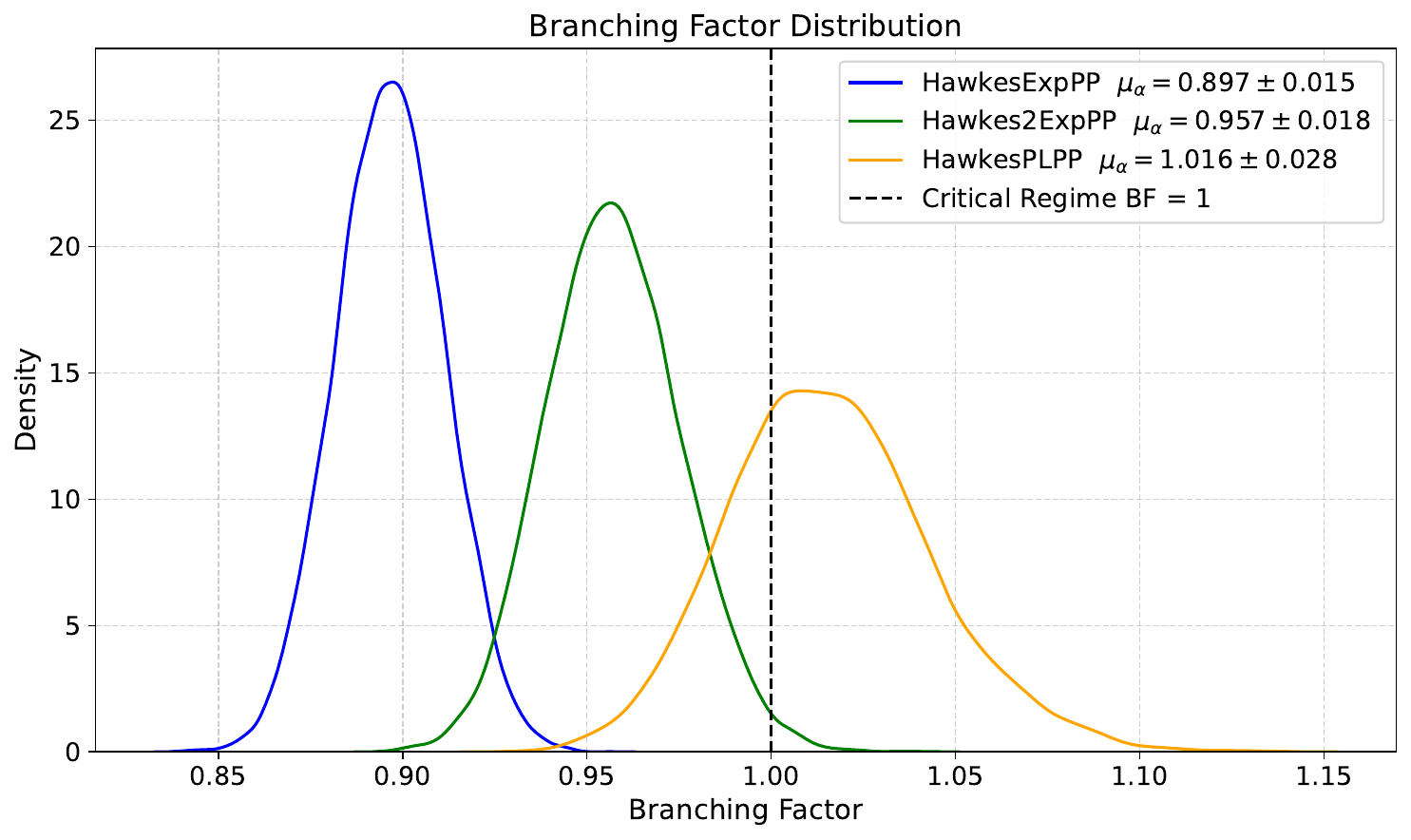}
    \caption{Posterior \gls{pdf} over the branching factor for the \glspl{hwkpp}.}
    \label{fig:bf}
\end{figure}

\begin{figure}[H]
    \centering
    \begin{minipage}{0.8\textwidth}
        \centering
        \includegraphics[width=\linewidth]{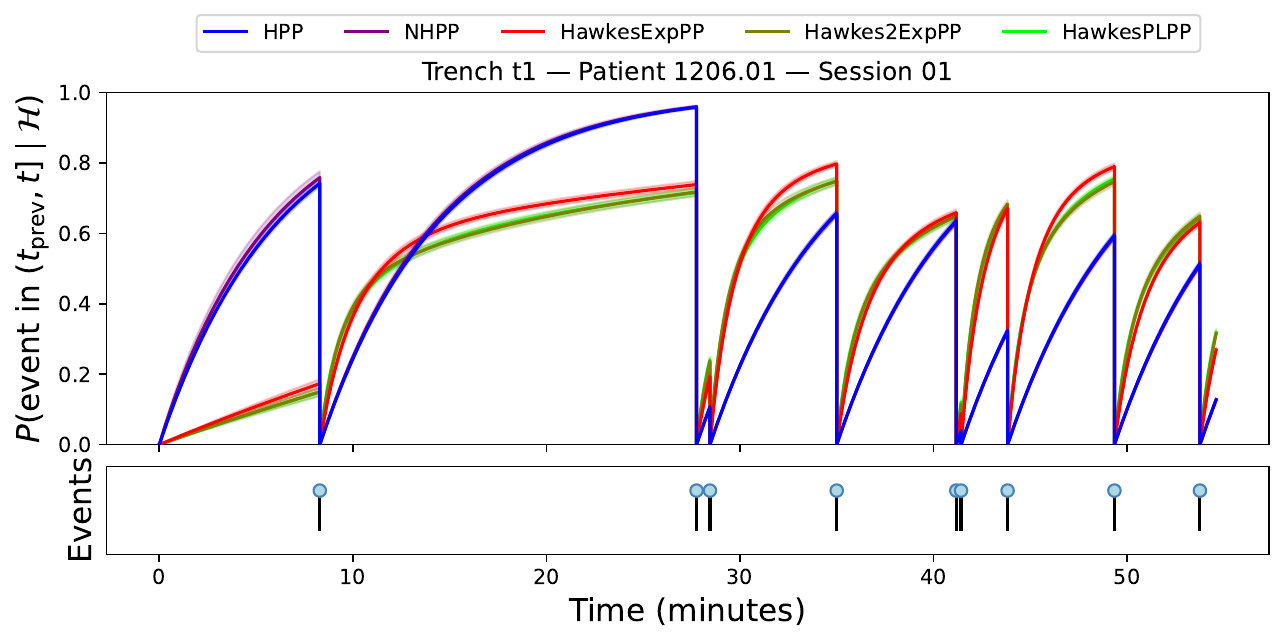}
        \subcaption{}
    \end{minipage}
    \begin{minipage}{0.8\textwidth}
        \centering
        \includegraphics[width=\linewidth]{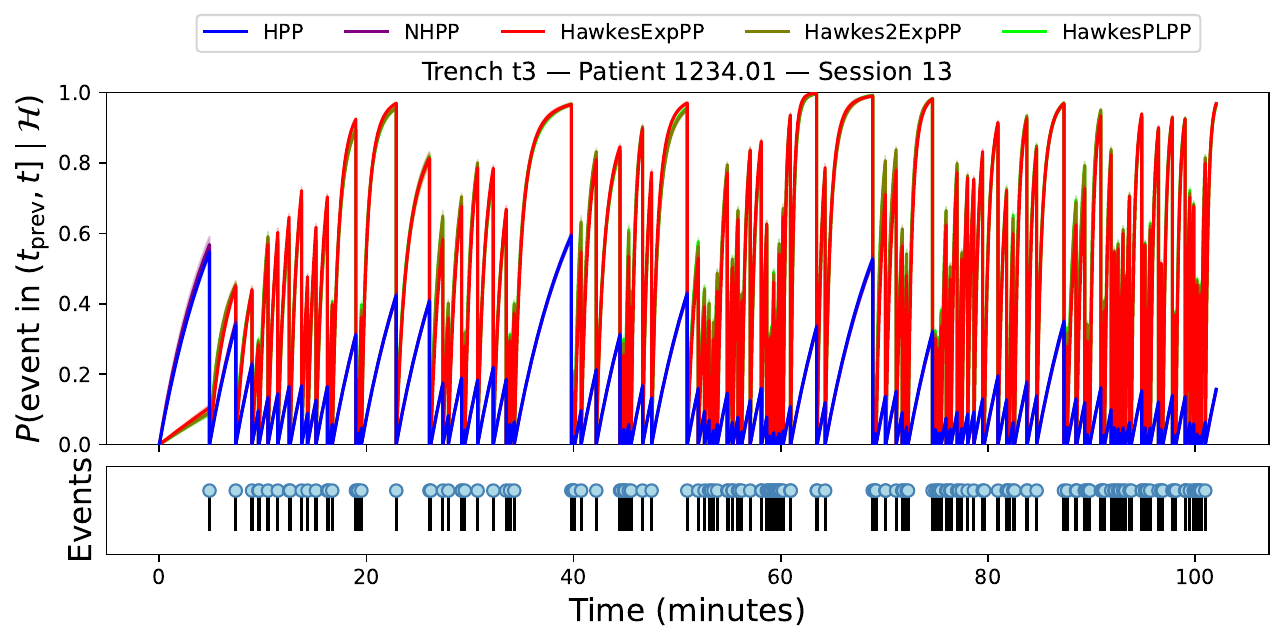}
        \subcaption{}
    \end{minipage}
    \begin{minipage}{0.8\textwidth}
        \centering
        \includegraphics[width=\linewidth]{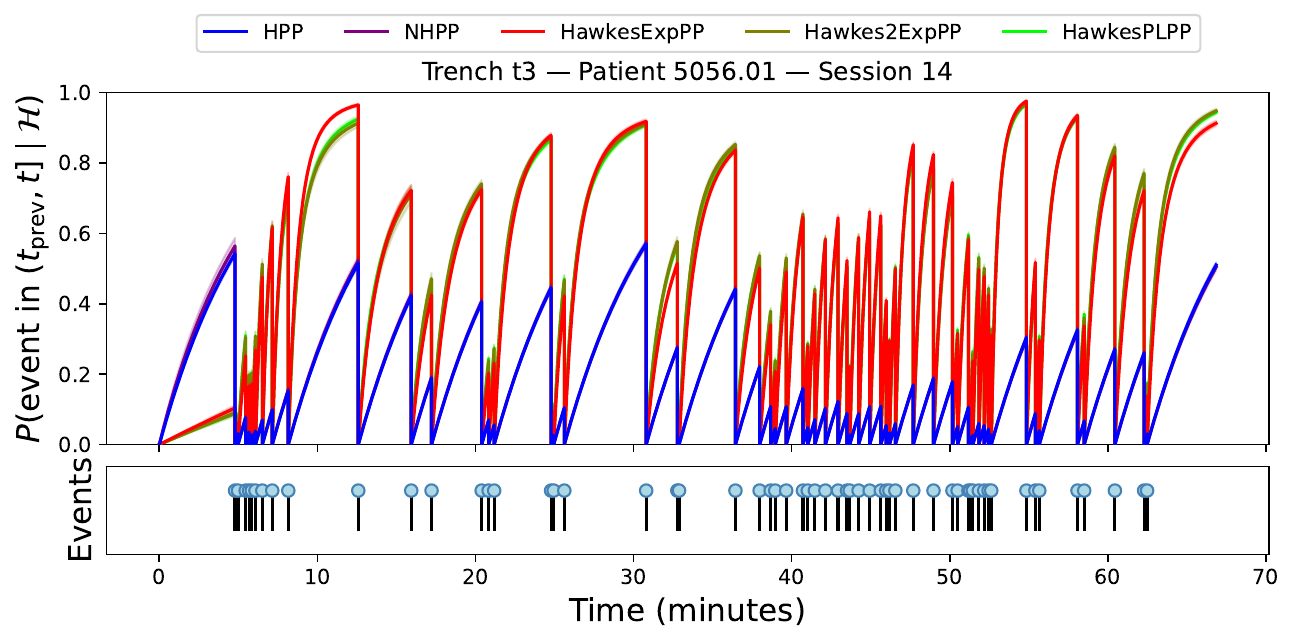}
        \subcaption{}
    \end{minipage}
    \caption{The median and 90\% credible interval of the \glspl{tpp} conditional \gls{cdf} curves for two observation sessions. The conditional \gls{cdf} curve of each \gls{tpp} model for (a) participant 1206.01 - observation session 1, (b) participant 1234.01 - observation session 13, (c) participant 5056.01 - observation session 14.}
    \label{fig:probability_fns}
\end{figure}

\begin{figure}[H]
    \centering
    \includegraphics[width=0.75\linewidth]{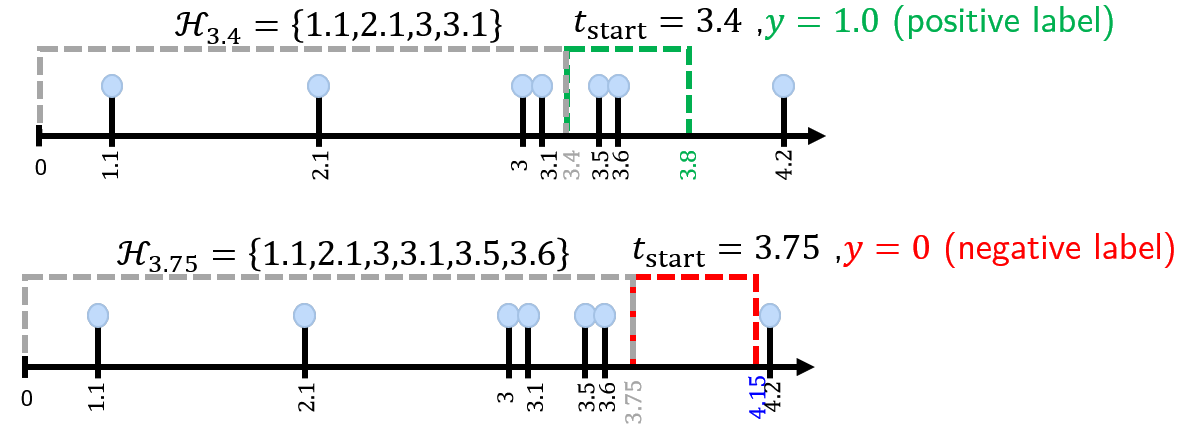}
    \caption{Two examples of sampling random windows and their corresponding history for the \gls{rocauc} evaluation. The black vertical lines mark onset timestamps. The blue horizontal line indicates the history $\h_{t_\text{start}}$ used by the \gls{tpp} for prediction; the green and red horizontal lines indicate positive and negative windows, respectively. A positive window (label 1) means at least one onset occurred in $[t_\text{start}, t_\text{start}+\Delta t]$; a negative window (label 0) means no onsets occurred.}
    \label{fig:example_windowing}
\end{figure}

\begin{table}[h!]
\centering
\scalebox{0.78}{
\begin{tabular}{|c|c|c|c|c|c|c|c|c|}
\hline
\textbf{$\Delta t$ (min)} & \textbf{\# of Samples} & \textbf{\# of + Samples} & \textbf{\# of - Samples} & \textbf{HPP} & \textbf{NHPP} & \textbf{HawkesExpPP} & \textbf{Hawkes2ExpPP} & \textbf{HawkesPLPP}\\ \hline
\textbf{1}        & 42900 & 4713   & 38187 & 0.501 & 0.541 & 0.848 & 0.852 & 0.848 \\
\textbf{5}        & 41700 & 10673  & 31027 & 0.499 & 0.546 & 0.824 & 0.826 & 0.827 \\
\textbf{10}       & 39900 & 13477  & 26423 & 0.490 & 0.536 & 0.806 & 0.805 & 0.806 \\
\textbf{15}       & 36400 & 14235  & 22165 & 0.496 & 0.531 & 0.802 & 0.800 & 0.802 \\
\textbf{20}       & 32800 & 14130  & 18670 & 0.501 & 0.525 & 0.777 & 0.781 & 0.783 \\
\textbf{25}       & 28700 & 13231  & 15469 & 0.494 & 0.512 & 0.778 & 0.783 & 0.785 \\ \hline
\end{tabular}
}
\caption{\gls{rocauc} for predicting whether at least one onset will occur in a time interval of size $\Delta t$ (min). The classification score is the median posterior probability that at least one onset occurs in $[t_\text{start}, t_\text{start} + \Delta t]$ across \gls{mcmc} samples. The number of samples is $|S|\times 100$, where $|S|$ varies with $\Delta t$ because a session is valid only if $T \geq 2\Delta t$. The number of positive and negative samples corresponds to ground-truth labels 1 and 0, respectively.}
\label{tab:rocauc}
\end{table}

\begin{table}[h!]
\centering
\begin{tabular}{|c|c|c|c|c|c|c|}
\toprule
Metric & |S| & HPP & NHPP & HawkesExpPP & Hawkes2ExpPP & HawkesPLPP \\
\midrule
WD $\pm$ SE & 429 & 9.67 ± 0.147 & 9.74 ± 0.131 & 2.28 ± 0.517 & 2.51 ± 0.711 & 2.51 ± 0.705 \\
$\text{PSIS-LOO ELPD}_\text{LOO} \pm$ SE& 429 & $-13727 \pm 1004$ & $-13733 \pm 1006$ & $-7808  \pm  482$ & $-7710 \pm 471$ & $-7700 \pm 470.27$ \\
$\text{MAPE}@\Delta t=1$   &  429  &  10.93 & 10.93 & 10.97 & 10.78 & 10.79 \\
$\text{MAPE}@\Delta t=5$   &  417  &  83.73 & 83.35 & 28.49 & 27.67 & 27.64 \\
$\text{MAPE}@\Delta t=10$  &  399  &  106.47 & 105.27 & 36.41 & 37.63 & 37.29 \\
$\text{MAPE}@\Delta t=15$  &  364  &  149.35 & 148.88 & 43.83 & 48.65 & 47.44 \\
$\text{MAPE}@\Delta t=20$   &  328  &  209.32 & 209.33 & 51.80 & 60.39 & 58.38 \\
$\text{MAPE}@\Delta t=25$   &  287  &  269.52 & 268.98 & 55.96 & 62.19 & 60.75 \\ 
\bottomrule
\end{tabular}
\caption{\gls{gof} and predictive metrics for \glspl{tpp}. $|S|$ is the number of observation sessions used to calculate each metric, and SE is the standard error. The first row is \gls{wd} (mean±std) between the empirical data count distribution versus the \gls{tpp} generated count distribution for the number of onsets within an observation session over 100 \gls{mc} trials. The second row is \gls{psis-loo} \gls{elpd} over $|S|=429$ observation sessions. The third row onward, i.e. any row with $\text{MAPE}@\Delta t=X$, is the \gls{mape} for forecasting the number of onsets in a time interval of size $\Delta t$ (min). The forecast is the median count from 250 sampled onset sequences (for each observation session).} 
\label{tab:metric} 
\end{table}

\begin{table}[h!]
\scalebox{0.8}{
\begin{tabular}{llllllllll}
\toprule
Statistic & Q90 & Q95 & Q99 & Max & N & Q95/Q90 & Q99/Q95 & Max/Q99 & outlier \\
Model &  &  &  &  &  &  &  &  &  \\
\midrule
Hawkes2ExpPP & 12.69 ± 12.09 & 19.37 ± 16.32 & 30.71 ± 21.21 & 35.99 ± 23.47 & 31.28 ± 44.40 & 1.89 ± 1.63 & 2.35 ± 2.57 & 1.40 ± 1.55 & 0.06 ± 0.24 \\
HawkesPLPP & 13.75 ± 12.53 & 21.14 ± 16.67 & 32.55 ± 21.14 & 37.10 ± 23.19 & 25.56 ± 30.87 & 1.86 ± 1.29 & 2.21 ± 2.45 & 1.24 ± 0.89 & 0.04 ± 0.21 \\
HawkesExpPP & 12.50 ± 12.52 & 19.66 ± 17.21 & 31.26 ± 21.79 & 36.48 ± 23.91 & 28.28 ± 36.39 & 1.91 ± 1.30 & 2.63 ± 3.20 & 1.40 ± 1.85 & 0.06 ± 0.26 \\
True Data & 9.86 ± 10.07 & 14.25 ± 13.01 & 20.49 ± 17.34 & 22.75 ± 18.92 & 29.13 ± 36.15 & 1.57 ± 0.51 & 1.70 ± 1.41 & 1.14 ± 0.26 & 0.01 ± 0.09 \\
\bottomrule
\end{tabular}
}
\caption{Comparison of summary statistics for data generated from the fitted \gls{tpp} versus the true data summary statistics. Q90, Q95, Q99 are the 90th, 95th, and 99th quantiles, whereas $N$ is the average number of onsets per observation session.}
\label{tab:branchingfactor}
\end{table}

\section{Legends}
\begin{itemize}
    \itemsep0em 
    \item \textbf{Figure 1}: Three stages of data preprocessing. (a) Aggregating \gls{sib},\gls{ato},\gls{ed} behavior labels into a condition label. (b) Preprocessing only the onset timestamps. (c) Normalizing the onset timestamps to the beginning of the observation session.
    \item \textbf{Figure 2}: The median and 90\% credible interval of the \glspl{tpp} conditional intensity curves for two observation sessions.  The conditional intensity curve for each \gls{tpp} model in (a) participant 1206.01 - observation session 1, (b) participant 1234.01 - observation session 13, and (c) participant 5056.01 - observation session 14.
    \item \textbf{Figure 3}: Goodness-of-fit evaluation metrics for the \glspl{tpp}. Subfigure (a): The empirical data count distribution (orange) versus \gls{tpp} generated count distribution (blue) for the number of onsets in an observation session. Subfigure (b): the \gls{qq} plot of \gls{rtc} theorem inter-arrivals for different \gls{tpp} fits. The x-axis is the theoretical quantiles of an exponential distribution with mean 1, and the y-axis is the sample quantiles from all the \gls{rtc} theorem inter-arrivals in the dataset.  Subfigure (c): the scatter plot of the raw residuals (\cref{eqn:rawresidual}) versus the number of onsets in an observation session (blue). In each subfigure, (a)-(e) refer to the \gls{hpp}, \gls{nhpp}, \gls{hwkexp}, \gls{hwk2exp}, and \gls{hwkpl} models, respectively.
    \item \textbf{Figure 4}: Visualization of the forecasted number of onsets within a future time window, conditioned on the history of prior onsets. The blue line represents the observed history of onset counts, while the black line denotes the true future onset count. The black vertical dotted line marks the start of forecasting. The red line corresponds to the median prediction over 250 sampled forecasts, with the darker orange shaded region indicating the interquartile range (25th to 75th quantile) and the lighter orange shaded region representing the 90\% credible interval (5th to 95th quantile). (a,b) show the forecasted number of onsets in participant 1206.01, session 3, (c,d) correspond to participant 4107.01, session 6, while (e,f) represent participant 5056.01, session 14, with each pair displaying forecasts for the \gls{hpp} and \gls{hwkexp}, respectively.
    \item \textbf{Figure 5}: Posterior \gls{pdf} over the branching factor for the \glspl{hwkpp}.
    \item \textbf{Figure 6}: The median and 90\% credible interval of the \glspl{tpp} conditional \gls{cdf} curves for two observation sessions. The conditional \gls{cdf} curve of each \gls{tpp} model for (a) participant 1206.01 - observation session 1, (b) participant 1234.01 - observation session 13, (c) participant 5056.01 - observation session 14.
    \item \textbf{Figure 7}: Two examples of sampling random windows and their corresponding history for the \gls{rocauc} evaluation. The black vertical lines mark onset timestamps. The blue horizontal line indicates the history used by the \gls{tpp} for prediction; the green and red horizontal lines indicate positive and negative windows, respectively. A positive window (label 1) means at least one onset occurred within the window; a negative window (label 0) means no onsets occurred.
    \item \textbf{Table 1}: \gls{rocauc} for predicting whether at least one onset will occur in a time interval of size $\Delta t$ (min). The classification score is the median posterior probability that at least one event occurs within the window across \gls{mcmc} samples. The number of samples is $|S|\times 100$, where $|S|$ varies with $\Delta t$ because a session is valid only if $T \geq 2\Delta t$. The number of positive and negative samples corresponds to ground-truth labels 1 and 0, respectively.
    \item \textbf{Table 2}: \gls{gof} and predictive metrics for \glspl{tpp}. $|S|$ is the number of observation sessions used to calculate each metric, and SE is the standard error. The first row is \gls{wd} (mean±std) between the empirical data count distribution versus the \gls{tpp} generated count distribution for the number of onsets within an observation session over 100 \gls{mc} trials. The second row is \gls{psis-loo} \gls{elpd} over $|S|=429$ observation sessions. The third row onward, i.e. any row with $\text{MAPE}@\Delta t=X$, is the \gls{mape} for forecasting the number of onsets in a time interval of size $\Delta t$ (min). The forecast is the median count from 250 sampled onset sequences (for each observation session).
    \item \textbf{Table 3}: Comparison of summary statistics for data generated from the fitted \gls{tpp} versus the true data summary statistics. Q90, Q95, Q99 are the 90th, 95th, and 99th quantiles, whereas $N$ is the average number of onsets per observation session.
\end{itemize}

\pagebreak \newpage

\resetlinenumber 
\linenumbers
\beginsupplement
\appendixpage
\glsresetall

\section{Supplementary Methods}

\subsection{Monte-Carlo Markov Chain Diagnostics}
\label{appsec:mcmcdiagnostics}

\begin{figure}[h!]
    \centering
    \includegraphics[width=0.99\linewidth]{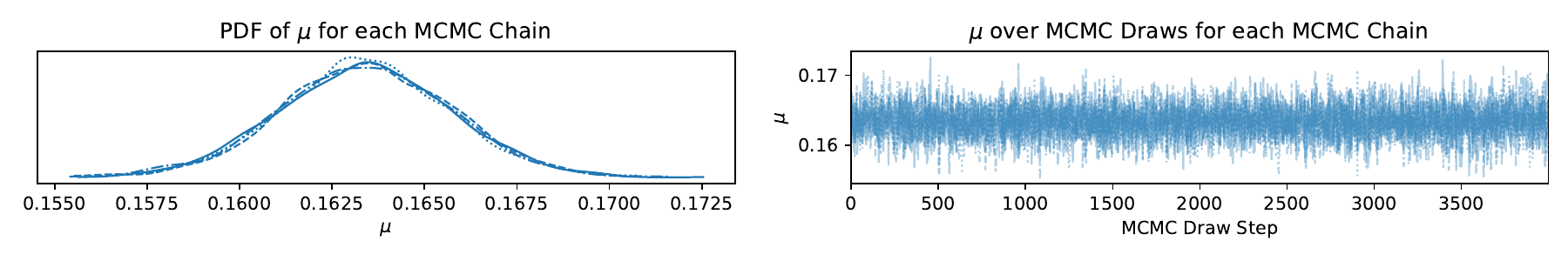}
    \caption{Trace plots for \gls{hpp}. Left column: Posterior probability density functions (PDFs) for each \gls{tpp} parameter, estimated from MCMC samples. Right column: Autocorrelation plots for each parameter's MCMC chain. Each subplot shows results from four independent MCMC chains, represented by four different lines.}
    \label{appfig:hppmcmchist}
\end{figure}

\begin{figure}[h!]
    \centering
    \includegraphics[width=0.99\linewidth]{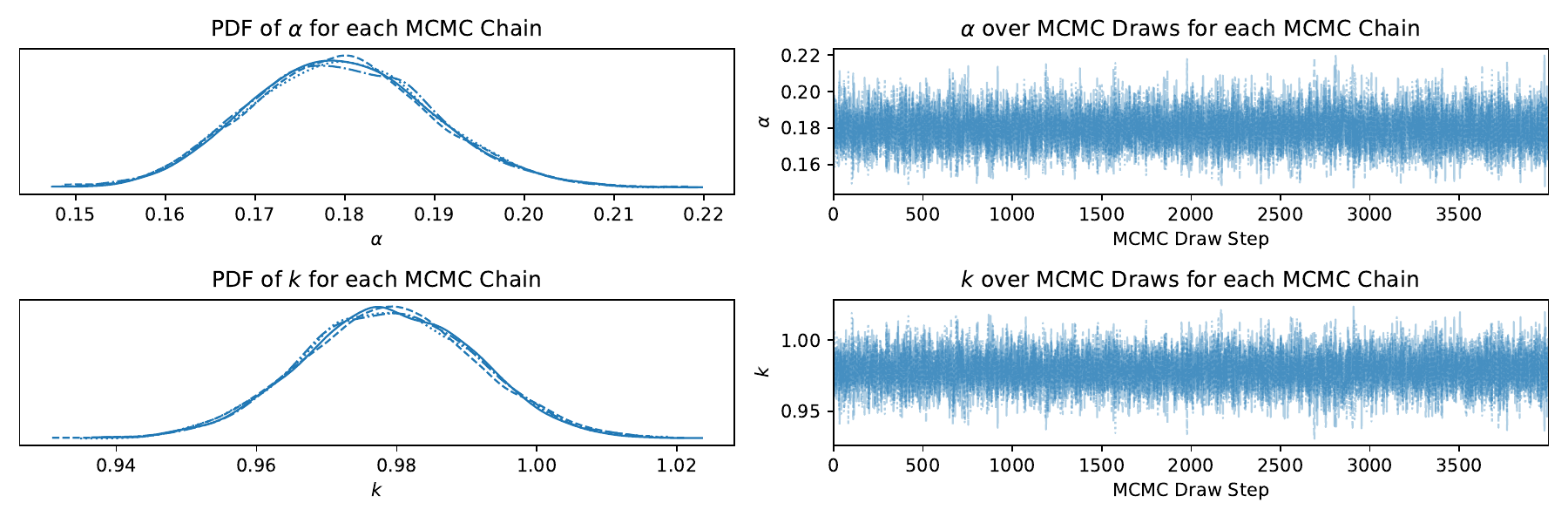}
    \caption{Trace plots for \gls{nhpp}. Left column: Posterior probability density functions (PDFs) for each \gls{tpp} parameter, estimated from MCMC samples. Right column: Autocorrelation plots for each parameter's MCMC chain. Each subplot shows results from four independent MCMC chains, represented by four different lines.}
    \label{appfig:nhppmcmchist}
\end{figure}

\begin{figure}[h!]
    \centering
    \includegraphics[width=0.99\linewidth]{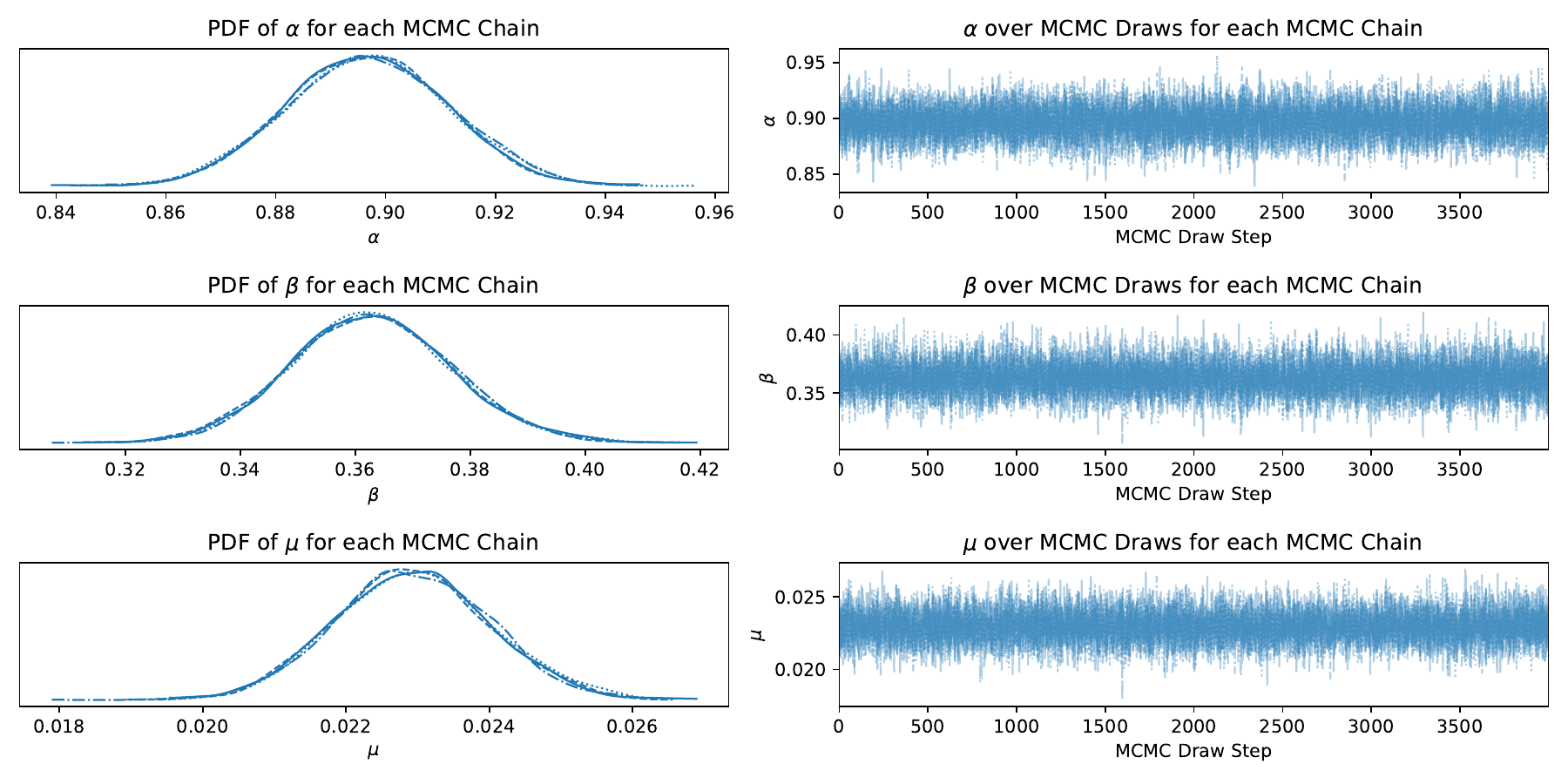}
    \caption{Trace plots for \gls{hwkexp}. Left column: Posterior probability density functions (PDFs) for each \gls{tpp} parameter, estimated from MCMC samples. Right column: Autocorrelation plots for each parameter's MCMC chain. Each subplot shows results from four independent MCMC chains, represented by four different lines.}   
    \label{appfig:hwkexpmcmchist}
\end{figure}

\begin{figure}[h!]
    \centering
    \includegraphics[width=0.99\linewidth]{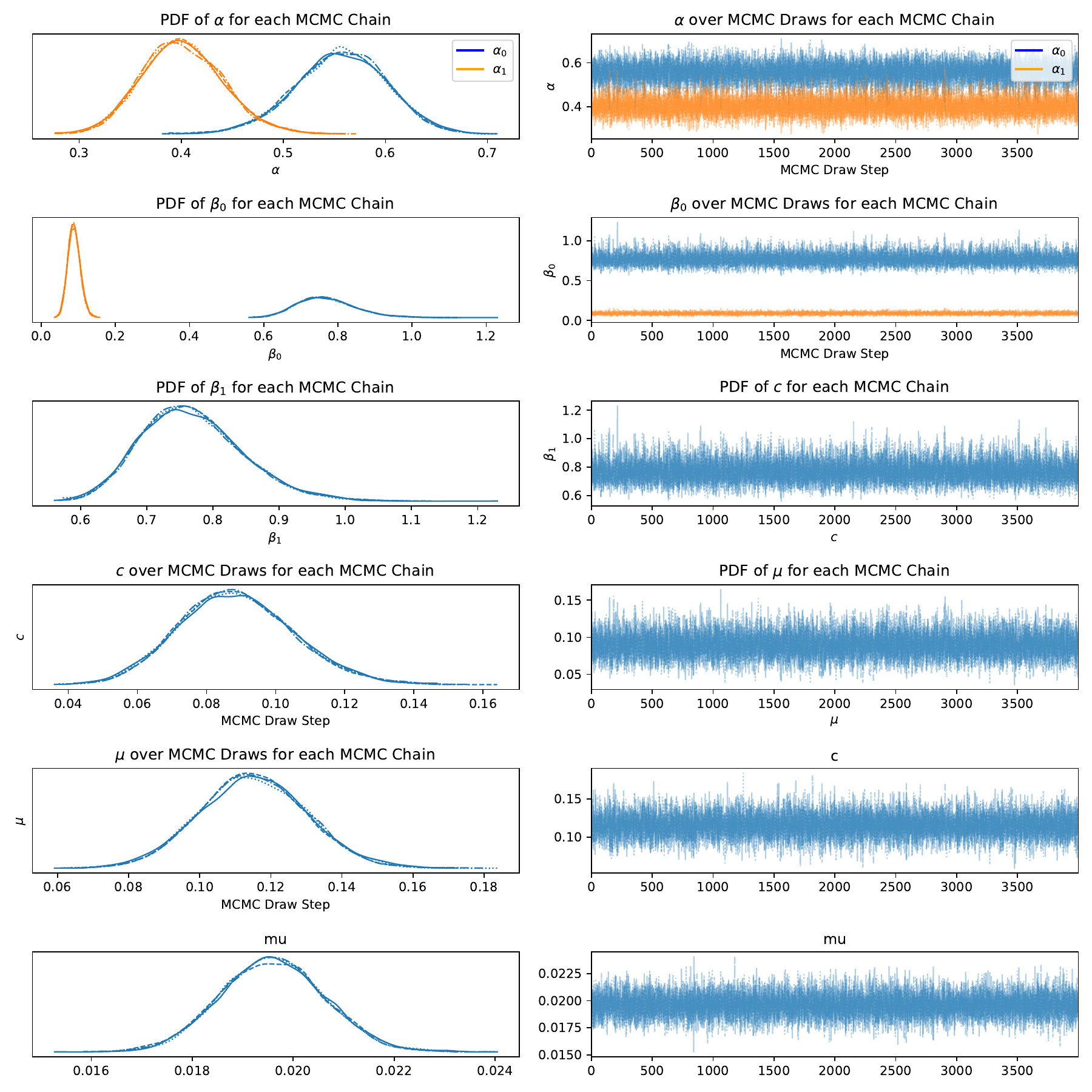}
    \caption{Trace plots for \gls{hwk2exp}. Left column: Posterior probability density functions (PDFs) for each \gls{tpp} parameter, estimated from MCMC samples. Right column: Autocorrelation plots for each parameter's MCMC chain. Each subplot shows results from four independent MCMC chains, represented by four different lines.}
        \label{appfig:hwk2expmcmchist}
\end{figure}

\begin{figure}[h!]
    \centering
    \includegraphics[width=0.99\linewidth]{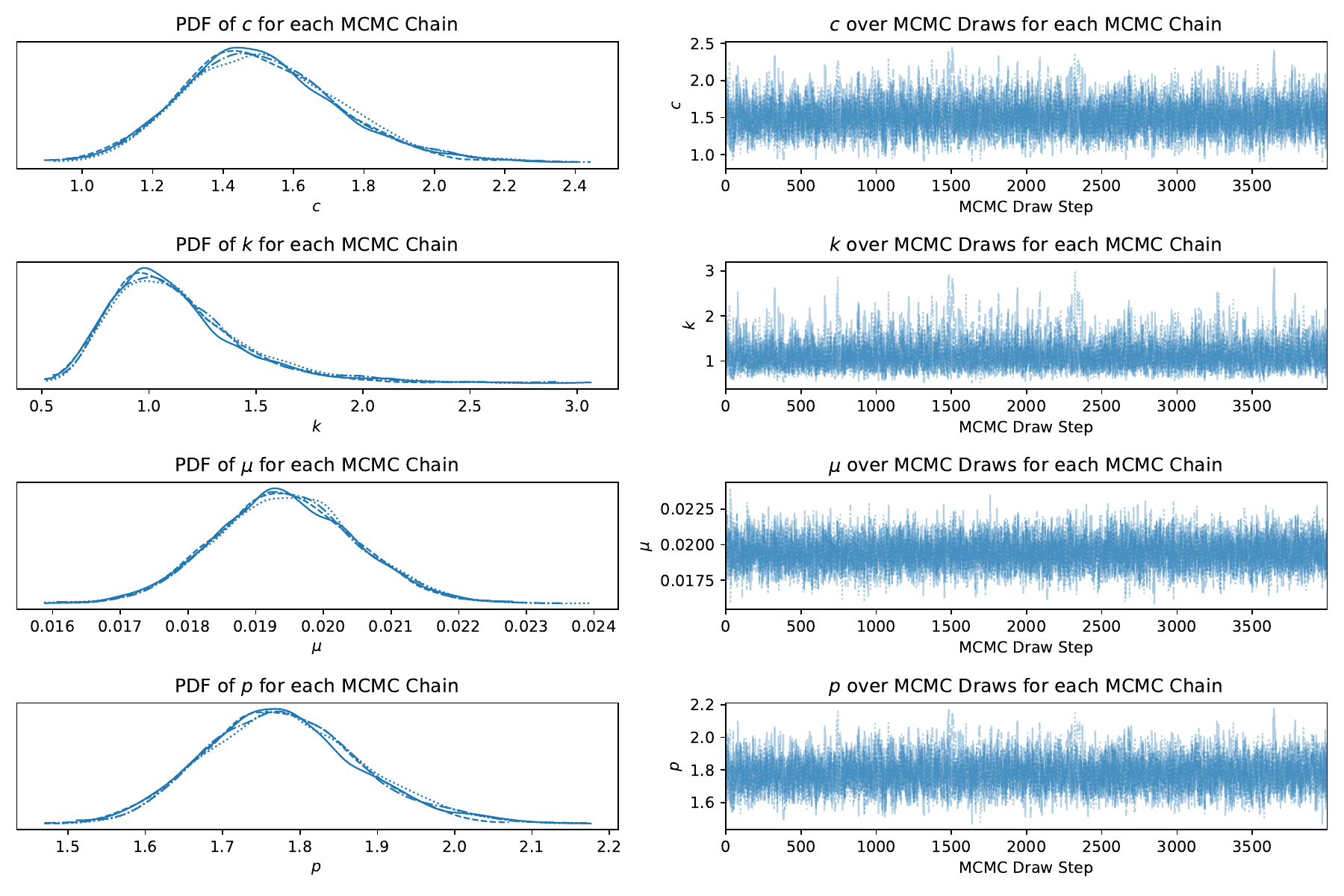}
    \caption{Trace plots for \gls{hwkpl}. Left column: Posterior probability density functions (PDFs) for each \gls{tpp} parameter, estimated from MCMC samples. Right column: Autocorrelation plots for each parameter's MCMC chain. Each subplot shows results from four independent MCMC chains, represented by four different lines.}
  \label{appfig:hwkplmcmchist}
\end{figure}

We provide the \gls{mcmc} inference diagnostics such as $\hat{R}$, \gls{mcse} on the posterior mean and posterior variance of the \gls{tpp} parameters, the \gls{ess} for the bulk and tail as well as the parameter expectations, standard deviations, and quantiles. Furthermore, we provide the autocorrelation plots of each \gls{mcmc} chain along with the posterior \gls{pdf} of the \gls{tpp} parameters. The \gls{mcmc} inference diagnostics and \gls{mcmc} sample summary statistics for the \gls{hpp}, \gls{nhpp}, \gls{hwkexp}, \gls{hwk2exp}, and \gls{hwkpl} are found in \Cref{apptab:hppmcmc,apptab:nhppmcmc,apptab:hwkexpmcmc,apptab:hwk2expmcmc,apptab:hwkplmcmc} respectively. The \gls{mcmc} sampling chain autocorrelation plots and \gls{mcmc} sample histograms for each parameter for \gls{hpp}, \gls{nhpp}, \gls{hwkexp}, \gls{hwk2exp}, and \gls{hwkpl} are shown in \cref{appfig:hppmcmchist,appfig:nhppmcmchist,appfig:hwkexpmcmchist,appfig:hwk2expmcmchist,appfig:hwkplmcmchist} respectively.

\subsection{\gls{psis-loo} \gls{elpd}}
\begin{figure}[h!]
    \centering
    \includegraphics[width=0.99\linewidth]{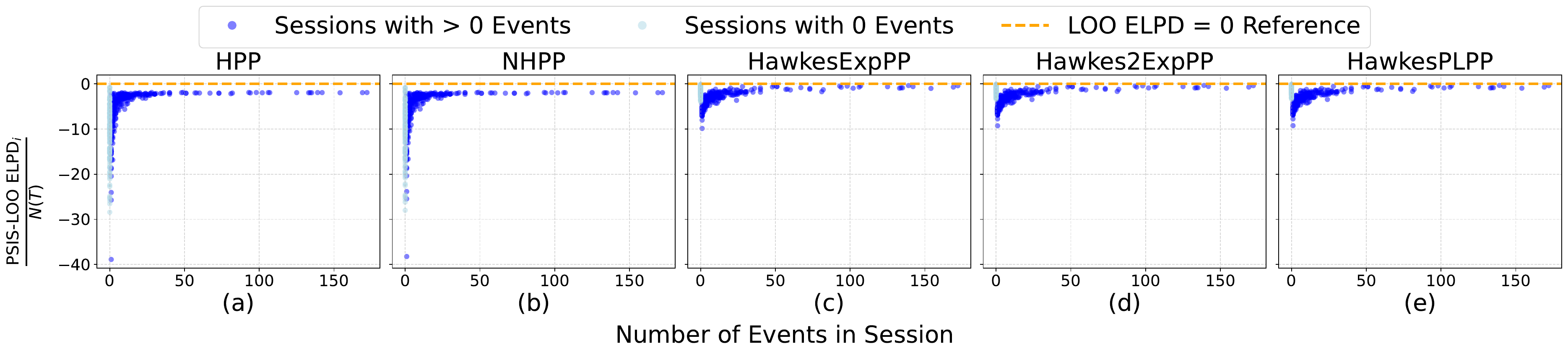}

    \caption{Visualization of the \gls{psis-loo} \gls{elpd} for each observation session normalized by number of onsets versus the number of onsets observed within an observation session. The light blue scatter points correspond to the \gls{psis-loo} \gls{elpd} for observation sessions with no onsets, whereas the darker blue scatter points correspond to \gls{psis-loo} \gls{elpd} for observation sessions with at least one onset. For sessions with no events, we simply divide the \gls{psis-loo} \gls{elpd} by 1. (a),(b),(c),(d),(e) correspond to the \gls{hpp}, \gls{nhpp}, \gls{hwkexp}, \gls{hwk2exp} and \gls{hwkpl} \glspl{tpp} respectively.}
    \label{appfig:elpd_scatterplot}
\end{figure}

We visualize the \gls{elpd} normalized by the number of aggressive behaviors in the observation session for each observation session against the number of observed aggressive behavior onsets to examine whether poor \gls{elpd} estimates correlate with the onset count (\cref{appfig:elpd_scatterplot}). Notably, the \gls{tpp} model performs worst when the number of onsets per session is between 1 and 10, indicating greater difficulty in capturing event dynamics in sparsely populated sessions.  

\subsection{Posterior versus Prior PDF of TPP parameters}
\begin{figure}[h!]
    \centering
    \includegraphics[width=0.7\linewidth]{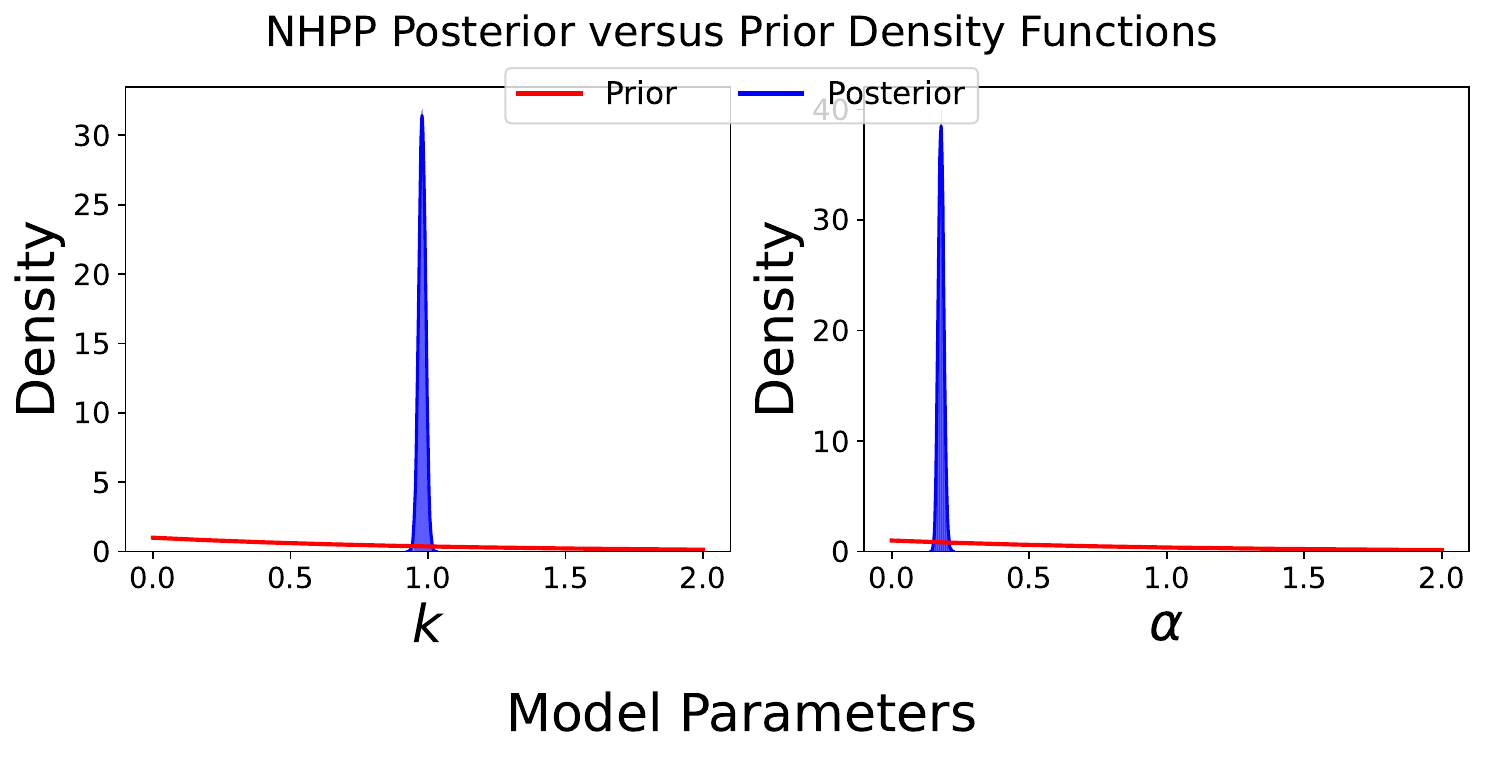}
    \caption{Posterior versus prior \gls{pdf} for \gls{nhpp} parameters. The prior distributions for the \gls{nhpp} parameters are: $k \sim \mathrm{Gamma}(1,1)$, $\alpha \sim \mathrm{Gamma}(1,1)$, and $\mu \sim \mathrm{Gamma}(1,1)$.}
    \label{appfig:priorpos_nhpp}
\end{figure}

\begin{figure}[h!]
    \centering
    \includegraphics[width=0.9\linewidth]{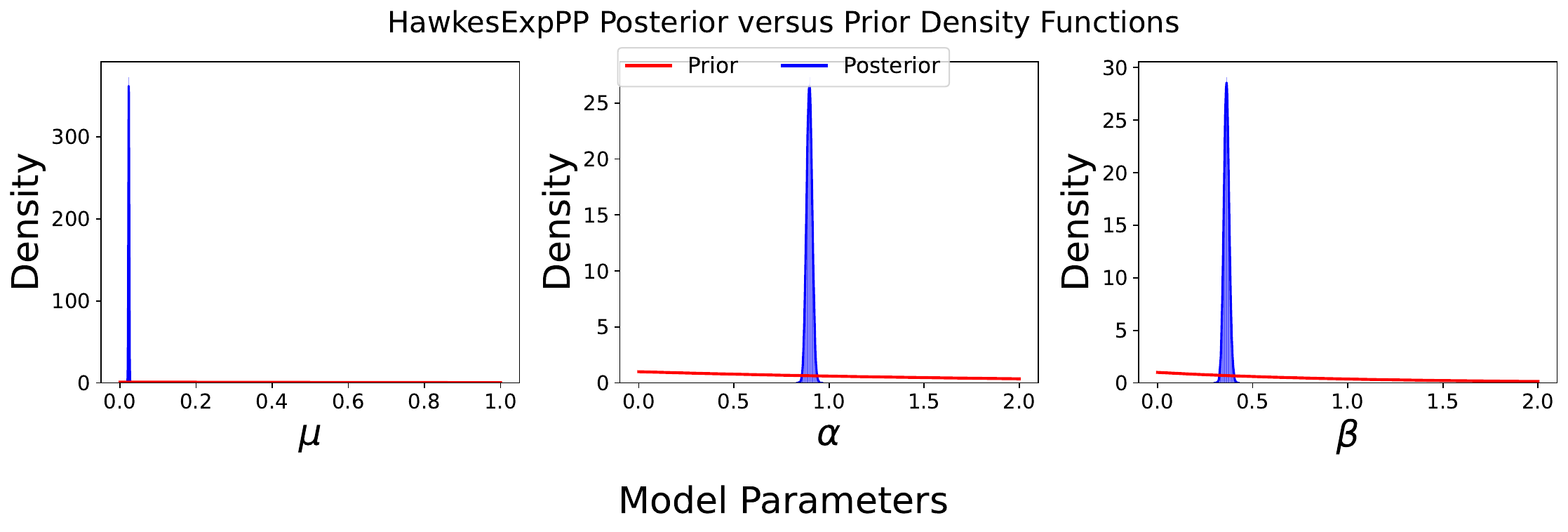}
    \caption{Posterior versus prior \gls{pdf} for \gls{hwkexp} parameters. The prior distributions for the \gls{hwkexp} parameters are: $\mu \sim \mathrm{Gamma}(1,1)$, $\alpha \sim \mathrm{Gamma}(1,1)$, and $\beta \sim \mathrm{Gamma}(1,1)$.}
    \label{appfig:priorpos_hwkexp}
\end{figure}

\begin{figure}[h!]
    \centering
    \includegraphics[width=0.7\linewidth]{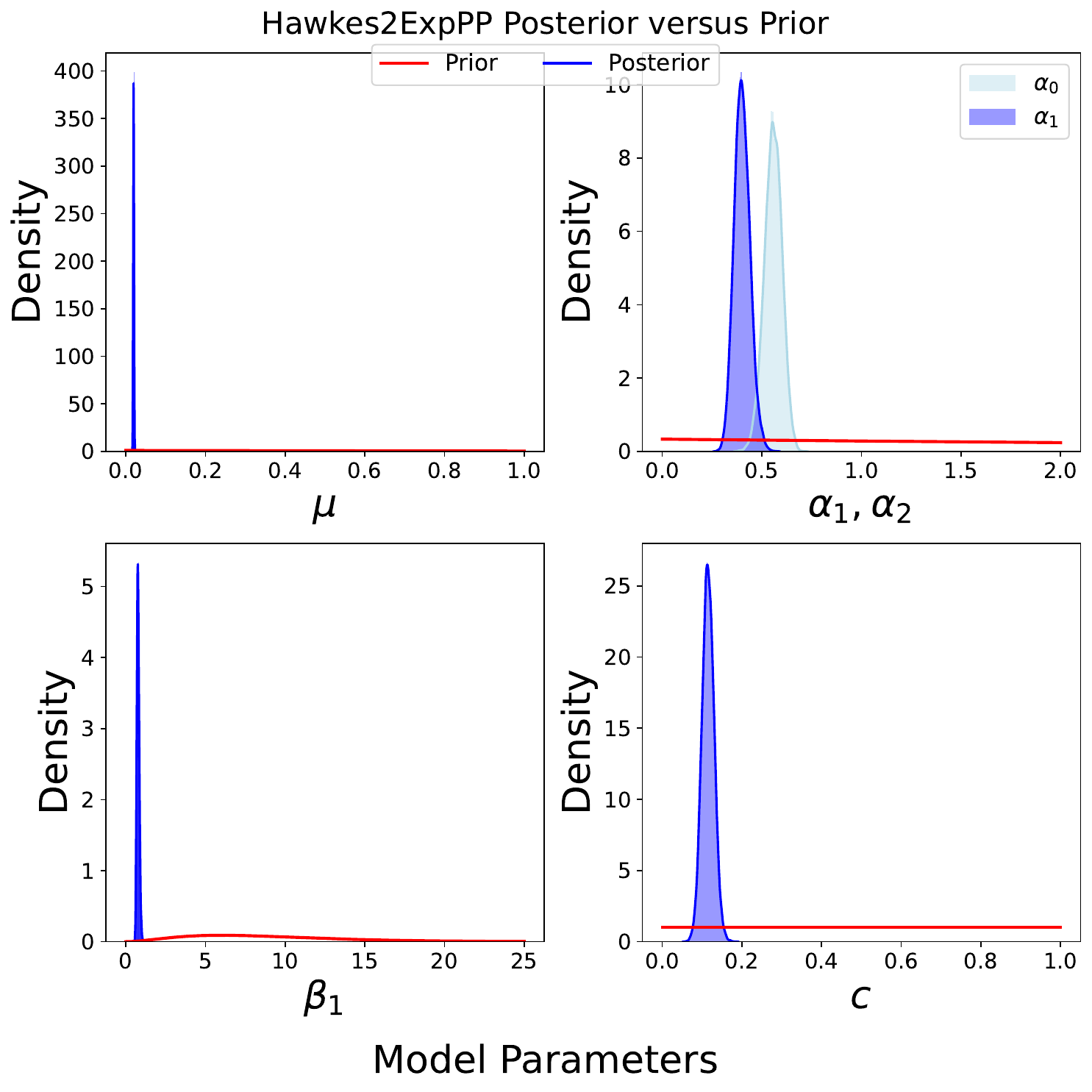}
    \caption{Posterior versus prior \gls{pdf} for \gls{hwk2exp} parameters. The prior distributions for the \gls{hwk2exp} parameters are: $\mu \sim \mathrm{Gamma}(1,1)$, $\alpha_0 \sim \mathrm{Gamma}(1,3)$, $\alpha_1 \sim \mathrm{Gamma}(1,3)$, $\beta_0 \sim \mathrm{Gamma}(3,3)$, $\beta_1 \sim \mathrm{Gamma}(3,3)$, and $c \sim \mathrm{Uniform}(0,1)$.}
    \label{appfig:priorpos_hwk2exp}
\end{figure}

\begin{figure}[h!]
    \centering
    \includegraphics[width=0.7\linewidth]{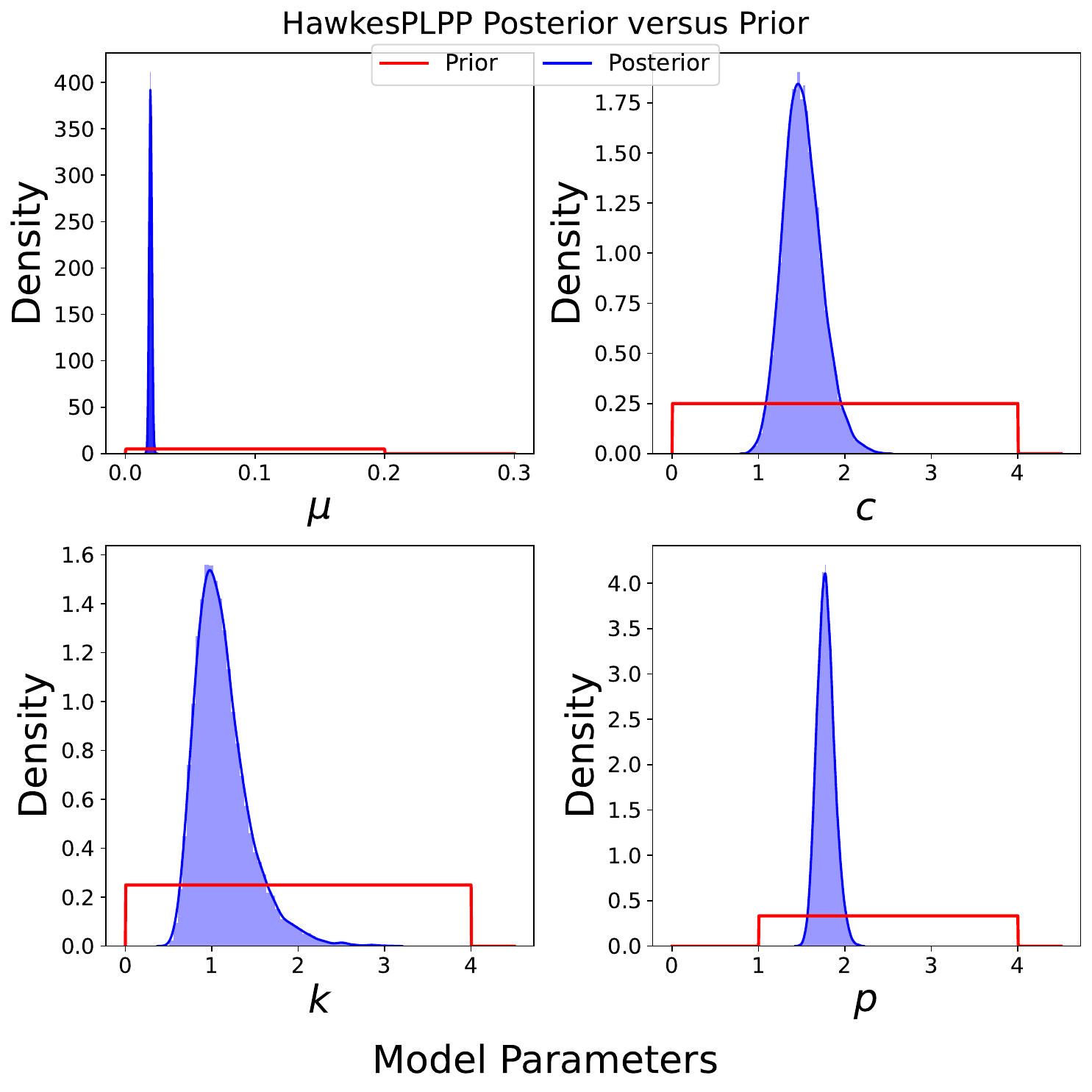}
    \caption{Posterior versus prior \gls{pdf} for \gls{hwkpl} parameters. The prior distributions for the \gls{hwkpl} parameters are: $\mu \sim \mathrm{Uniform}(0,2)$, $c \sim \mathrm{Uniform}(0,4)$, $k \sim \mathrm{Uniform}(0,4)$, and $p \sim \mathrm{Uniform}(1,4)$.}
    \label{appfig:priorpos_hwkpl}
\end{figure}

We visualize the posterior \gls{pdf} from \gls{mcmc} inference samples and the prior \gls{pdf} for each model parameter to compare how much the data shifts the posterior \gls{pdf} from the prior \gls{pdf}. This gives some indication on how much the data drives the posterior \gls{pdf}, and if the prior distribution is biasing the results. The red lines and blue lines in the figures correspond to the prior \gls{pdf} and posterior \gls{pdf} respectively on the \gls{tpp} parameters. As shown in \cref{appfig:priorpos_nhpp,appfig:priorpos_hwkexp,appfig:priorpos_hwk2exp,appfig:priorpos_hwkpl}, with the exception of the \gls{hwkpl}, the prior \glspl{pdf} are widely uninformative priors and the data drives the posterior \gls{pdf}. In future work, we will try wider prior \glspl{pdf} for the \gls{hwkpl}.

\subsection{Ripley K Estimator}
\label{appsec:ripley}

\begin{figure}[h!]
    \centering
    \includegraphics[trim={100pt 0pt 100pt 0pt},clip,width=0.8\linewidth]{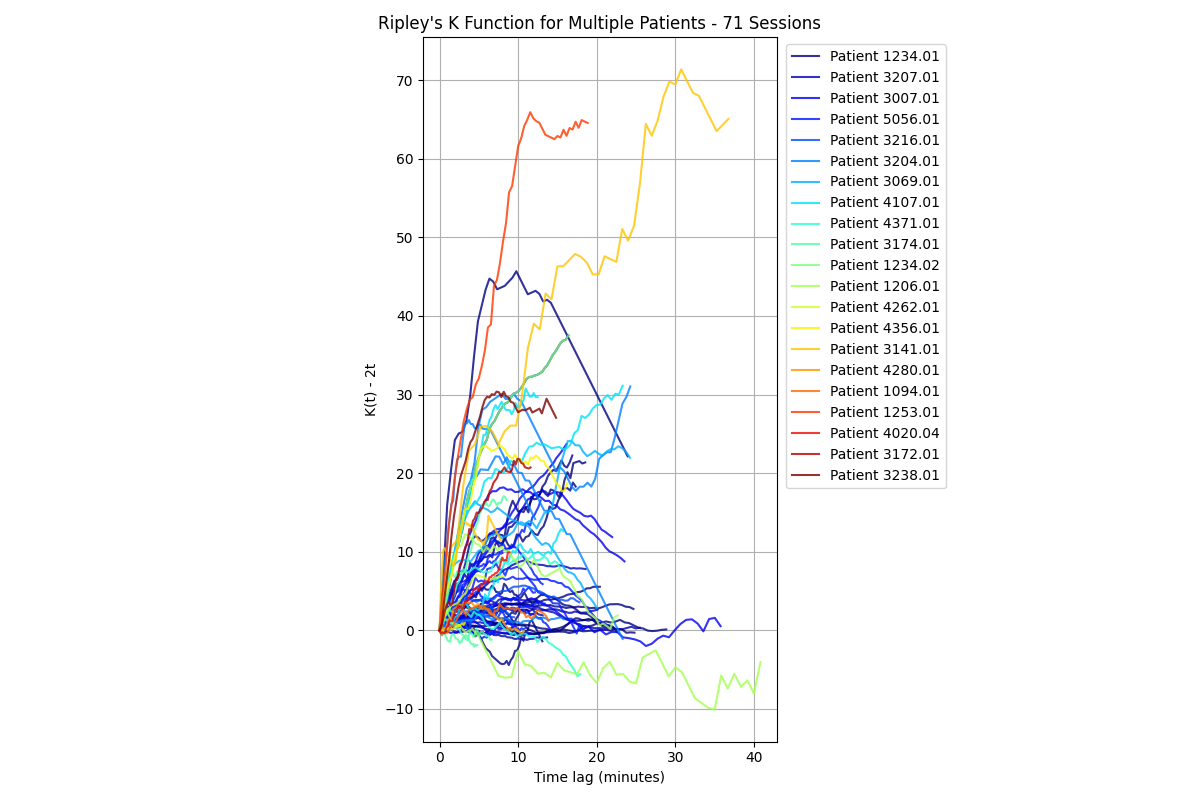}
    \caption{For each observation session with at least 15 onsets, we visualize the Ripley K estimator for varying values of $t$ with reference to the \gls{hpp} Ripley K estimator $2t$. The time lag is $t$.}
    \label{appfig:ripleyk}
\end{figure}

For self-exciting point processes, each onset triggers other onsets with high probability, but it is not necessarily true that the onset times cluster temporally. A method to investigate the aggregation / clustering of onsets is to use the non-parametric Ripley's K-function summary \cite{karlis2023proposer}, which is a reduced second-moment measure. It may be thought of as a normalized (with respect to the rate of events) version of the expected number of onsets within a window of interval $2t$ of any random onset time selected (\Cref{appeq:ripleyk}). For a \gls{hpp}, theoretically $K(t)=2t$. If $K(t)>2t$, the process is said to be over-dispersed relative to the \gls{hpp} and thus exhibits some degree of clustering. If $K(t)<2t$, the process is said to be under-dispersed related to the \gls{hpp} and thus onsets occur towards regular intervals. As shown in \cref{appfig:ripleyk}, many of the observation sessions display some form of temporal clustering, indicating that self-exciting processes are an appropriate model choice.

\subsection{Time Complexity}
\label{appsec:time_complexity}
We provide a detailed analysis covering both theoretical time complexity and empirical runtime (\cref{apptab:intensity_complexity,apptab:timing_results}) for computing the conditional intensity (for an individual event, and over all events in an observation session). \Cref{apptab:intensity_complexity} summarizes the Big-$O$ complexity of evaluating the conditional intensity for a single new event and for all $N$ events in a session~\cite{laub2024hawkes}. Notably, the exponential kernel admits a recursive formulation that reduces per-event cost from $O(N)$ to $O(1)$, yielding $O(N)$ per session-matching the non-history-dependent baselines. To complement the theoretical analysis, \cref{apptab:timing_results} reports empirical runtime averaged over 10 Monte Carlo trials across 285 sessions (each containing at least one event) and 4{,}871 total events, measured on a 12th Gen Intel Core i-series processor at 2.5\,GHz under typical laptop load. All Hawkes variants evaluate the intensity for a full session in under 25\,ms, confirming that the computational overhead is negligible for practical use. These runtime results (\cref{apptab:timing_results}) were run on a 12th Gen Intel(R) Core(TM) at 2.5 GHz, with background processes running.

\newpage

\section{Supplementary Tables}

\begin{table}[h!]
\scalebox{0.8}{
\begin{tabular}{l|l|l|l|l|l|l|l|l|l|l|l|}
\hline
\multicolumn{1}{|c|}{\textbf{$\theta$}} & \multicolumn{1}{c|}{mcse\_mean} & \multicolumn{1}{c|}{mcse\_sd} & ess\_bulk & ess\_tail   & r\_hat   & mean     & std      & median\_std & 5\%      & median   & 95\%     \\ \hline
\multicolumn{1}{|l|}{$\mu$}      & 0.000039 & 0.000031 & 3633.801169 & 3656.278866 & 1.000362 & 0.163394 & 0.002333 & 0.002333 & 0.159559 & 0.163389 & 0.167252 \\ \hline
\end{tabular}}
\caption{\gls{hpp} \gls{tpp} \gls{mcmc} diagnostics. Table includes the \gls{mcmc} diagnostic statistics and the summary statistics over \gls{mcmc} samples of the model parameters.}
\label{apptab:hppmcmc}
\end{table}

\begin{table}[h!]
\scalebox{0.8}{
\begin{tabular}{|l|l|l|l|l|l|l|l|l|l|l|l|}
\hline
\multicolumn{1}{|c|}{\textbf{$\theta$}} & \multicolumn{1}{c|}{mcse\_mean} & \multicolumn{1}{c|}{mcse\_sd} & ess\_bulk & ess\_tail   & r\_hat   & mean     & std      & median\_std & 5\%      & median   & 95\%     \\ \hline
$\alpha$                           & 0.000112                                 & 0.000094                               & 8179.859153                             & 7653.780288                    & 1.000160                    & 0.179535                  & 0.010042                 & 0.010048                         & 0.163490                 & 0.179137                    & 0.196577                  \\ 
k                               & 0.000138                                 & 0.000115                               & 8202.293244                             & 7482.146427                    & 1.000065                    & 0.978895                  & 0.012380                 & 0.012380                         & 0.958507                 & 0.978955                    & 0.999088                  \\ 
$\mu$                             & 0.000070                                 & 0.000058                               & 8185.835066                             & 7844.571412                    & 1.000081                    & 0.172916                  & 0.002333                 & 0.002333                         & 0.159559                 & 0.163389                    & 0.167252                  \\ \hline
\end{tabular}
}
\caption{\gls{nhpp} \gls{tpp} \gls{mcmc} diagnostics. Table includes the \gls{mcmc} inference diagnostic statistics and the summary statistics over \gls{mcmc} samples of the model parameters.}
\label{apptab:nhppmcmc}
\end{table}
\vspace{-1em}

\begin{table}[h!]
\scalebox{0.8}{
\begin{tabular}{|l|l|l|l|l|l|l|l|l|l|l|l|}
\hline
\multicolumn{1}{|c|}{\textbf{$\theta$}} & \multicolumn{1}{c|}{mcse\_mean} & \multicolumn{1}{c|}{mcse\_sd} & ess\_bulk & ess\_tail   & r\_hat   & mean     & std      & median\_std & 5\%      & median   & 95\%     \\ \hline
$\alpha$ & 0.000135 & 0.000120 & 11829.301438 & 10332.547806 & 1.000292 & 0.896698 & 0.014693 & 0.014693 & 0.872563 & 0.896660 & 0.920973 \\
$\beta$ & 0.000130 & 0.000125 & 11637.630262 & 9383.983263 & 1.000407 & 0.362831 & 0.013994 & 0.013996 & 0.340169 & 0.362598 & 0.386296 \\
$\mu$ & 0.000010 & 0.000009 & 11349.449156 & 9683.741236 & 1.000642 & 0.022933 & 0.001083 & 0.001083 & 0.021177 & 0.022917 & 0.024750 \\ \hline
\end{tabular}
}
\caption{\gls{hwkexp} \gls{tpp} \gls{mcmc} diagnostics. Table includes the \gls{mcmc} inference diagnostic statistics and the summary statistics over \gls{mcmc} samples of the model parameters.}
\label{apptab:hwkexpmcmc}
\end{table}
\vspace{-1em}

\begin{table}[h!]
\scalebox{0.8}{
\begin{tabular}{|l|l|l|l|l|l|l|l|l|l|l|l|}
\hline
\multicolumn{1}{|c|}{\textbf{$\theta$}} & \multicolumn{1}{c|}{mcse\_mean} & \multicolumn{1}{c|}{mcse\_sd} & ess\_bulk & ess\_tail   & r\_hat   & mean     & std      & median\_std & 5\%      & median   & 95\%     \\ \hline
$\alpha_0$ & 0.000433 & 0.000366 & 10616.930211 & 8656.655341 & 1.000281 & 0.556320 & 0.044248 & 0.044258 & 0.481201 & 0.557267 & 0.626812 \\
$\alpha_1$ & 0.000379 & 0.000314 & 10712.089546 & 9802.366819 & 1.000084 & 0.400971 & 0.039333 & 0.039367 & 0.338864 & 0.399334 & 0.467710 \\
$\beta_0$ & 0.000743 & 0.000679 & 11043.649204 & 9629.683342 & 0.999968 & 0.768517 & 0.076621 & 0.076869 & 0.655544 & 0.762342 & 0.903744 \\
$\beta_1$ & 0.000157 & 0.000133 & 10771.045821 & 8949.318270 & 1.000410 & 0.088716 & 0.016540 & 0.016553 & 0.062614 & 0.088057 & 0.117224 \\
$c$ & 0.000132 & 0.000129 & 13019.508176 & 8931.120605 & 1.000558 & 0.114967 & 0.015083 & 0.015084 & 0.090576 & 0.114746 & 0.139882 \\
$\mu$ & 0.000008 & 0.000009 & 16403.539754 & 11628.560377 & 1.000222 & 0.019588 & 0.001028 & 0.001028 & 0.017920 & 0.019577 & 0.021295  \\ \hline
\end{tabular}
}
\caption{\gls{hwk2exp} \gls{tpp} \gls{mcmc} diagnostics. Table includes the \gls{mcmc} inference diagnostic statistics and the summary statistics over \gls{mcmc} samples of the model parameters.}
\label{apptab:hwk2expmcmc}
\end{table}
\vspace{-1em}

\begin{table}[h!]
\scalebox{0.8}{
\begin{tabular}{|l|l|l|l|l|l|l|l|l|l|l|l|}
\hline
\multicolumn{1}{|c|}{\textbf{$\theta$}} & \multicolumn{1}{c|}{mcse\_mean} & \multicolumn{1}{c|}{mcse\_sd} & ess\_bulk & ess\_tail   & r\_hat   & mean     & std      & median\_std & 5\%      & median   & 95\%     \\ \hline
$c$ & 0.004230 & 0.003565 & 3005.419284 & 2480.668995 & 1.001898 & 1.510057 & 0.222294 & 0.222911 & 1.172446 & 1.493486 & 1.901068 \\
$k$ & 0.006658 & 0.008944 & 2879.622558 & 2336.711490 & 1.001861 & 1.124569 & 0.314497 & 0.319398 & 0.725962 & 1.068830 & 1.718204 \\
$\mu$ & 0.000011 & 0.000010 & 8182.946713 & 8031.414127 & 1.000369 & 0.019448 & 0.001034 & 0.001034 & 0.017774 & 0.019425 & 0.021183 \\
$p$ & 0.001907 & 0.001466 & 2919.065331 & 2440.831268 & 1.001747 & 1.778094 & 0.099775 & 0.099885 & 1.623239 & 1.773396 & 1.953642 \\ \hline
\end{tabular}
}
\caption{\gls{hwkpl} \gls{tpp} \gls{mcmc} diagnostics. Table includes the \gls{mcmc} inference diagnostic statistics and the summary statistics over \gls{mcmc} samples of the model parameters. Diverging samples: 0.}
\label{apptab:hwkplmcmc}
\end{table}
\vspace{-1em}

\begin{table}[h!]
\centering
\scalebox{0.7}{
\begin{tabular}{|c|c|c|c|c|}
\hline
Patient ID & \# Of Onsets & \# of Observation Sessions & Total Observation Time (min) & Time (min)/ Session \\ \hline
1234.01    & 1287         & 24                         & 2000.08                      & 83.34               \\
3207.01    & 741          & 7                          & 528.88                       & 75.55               \\
3007.01    & 481          & 7                          & 629.65                       & 89.95               \\
5056.01    & 407          & 17                         & 940.49                       & 55.32               \\
3216.01    & 195          & 7                          & 434.00                       & 62.00               \\
4107.01    & 150          & 16                         & 1052.95                      & 65.81               \\
3204.01    & 141          & 9                          & 931.94                       & 103.55              \\
3069.01    & 118          & 3                          & 262.33                       & 87.44               \\
3174.01    & 98           & 8                          & 376.91                       & 47.11               \\
4020.04    & 94           & 16                         & 719.15                       & 44.95               \\
4371.01    & 79           & 5                          & 339.70                       & 67.94               \\
4262.01    & 79           & 7                          & 211.61                       & 30.23               \\
1094.01    & 77           & 25                         & 2263.22                      & 90.53               \\
1234.02    & 65           & 3                          & 253.31                       & 84.44               \\
4358.01    & 64           & 20                         & 746.63                       & 37.33               \\
1206.01    & 57           & 4                          & 386.73                       & 96.68               \\
1224.01    & 55           & 6                          & 740.04                       & 123.34              \\
4356.01    & 55           & 4                          & 177.53                       & 44.38               \\
4280.01    & 54           & 5                          & 203.55                       & 40.71               \\
1247.01    & 50           & 20                         & 1597.28                      & 79.86               \\
3141.01    & 44           & 5                          & 446.34                       & 89.27               \\
3172.01    & 42           & 7                          & 377.38                       & 53.91               \\
1229.01    & 38           & 11                         & 949.90                       & 86.35               \\
1236.01    & 37           & 13                         & 987.37                       & 75.95               \\
4348.01    & 33           & 7                          & 550.03                       & 78.58               \\
1235.01    & 33           & 9                          & 1088.57                      & 120.95              \\
4343.01    & 28           & 4                          & 164.43                       & 41.11               \\
4293.04    & 26           & 13                         & 792.47                       & 60.96               \\
1253.01    & 22           & 5                          & 558.22                       & 111.64              \\
4365.01    & 22           & 9                          & 330.09                       & 36.68               \\
3171.01    & 21           & 6                          & 267.56                       & 44.59               \\
1241.01    & 21           & 8                          & 657.87                       & 82.23               \\
3238.01    & 19           & 3                          & 204.75                       & 68.25               \\
4407.01    & 19           & 4                          & 296.57                       & 74.14               \\
1205.01    & 14           & 7                          & 772.40                       & 110.34              \\
4354.01    & 13           & 7                          & 516.60                       & 73.80               \\
1175.01    & 12           & 6                          & 384.54                       & 64.09               \\
4401.01    & 10           & 6                          & 256.91                       & 42.82               \\
4409.01    & 9            & 1                          & 77.82                        & 77.82               \\
1233.01    & 9            & 18                         & 908.44                       & 50.47               \\
4384.01    & 9            & 6                          & 278.54                       & 46.42               \\
1223.01    & 8            & 5                          & 686.19                       & 137.24              \\
1181.01    & 6            & 10                         & 644.56                       & 64.46               \\
4183.01    & 6            & 6                          & 252.44                       & 42.07               \\
4353.01    & 5            & 5                          & 241.64                       & 48.33               \\
4090.01    & 5            & 1                          & 75.77                        & 75.77               \\
4398.01    & 4            & 1                          & 71.95                        & 71.95               \\
2036.01    & 3            & 1                          & 22.32                        & 22.32               \\
4342.01    & 2            & 2                          & 129.08                       & 64.54               \\
4381.01    & 2            & 3                          & 196.33                       & 65.44               \\
4380.01    & 1            & 2                          & 103.88                       & 51.94               \\
4395.01    & 1            & 2                          & 80.94                        & 40.47               \\
1220.01    & 1            & 6                          & 762.71                       & 127.12              \\
1244.01    & 0            & 1                          & 4.98                         & 4.98                \\
1238.01    & 0            & 1                          & 112.40                       & 112.40              \\
1240.01    & 0            & 1                          & 64.95                        & 64.95               \\
1242.01    & 0            & 1                          & 49.75                        & 49.75               \\
4392.01    & 0            & 1                          & 66.48                        & 66.48               \\
4391.01    & 0            & 1                          & 40.00                        & 40.00               \\
4389.01    & 0            & 1                          & 26.78                        & 26.78               \\
4169.01    & 0            & 1                          & 46.63                        & 46.63               \\
1245.01    & 0            & 1                          & 73.85                        & 73.85               \\
1249.01    & 0            & 1                          & 27.18                        & 27.18               \\
1255.01    & 0            & 1                          & 71.05                        & 71.05               \\
4357.01    & 0            & 1                          & 45.85                        & 45.85               \\
1232.01    & 0            & 1                          & 79.50                        & 79.50               \\
1230.01    & 0            & 1                          & 39.83                        & 39.83               \\
3177.01    & 0            & 1                          & 67.88                        & 67.88               \\
3237.01    & 0            & 1                          & 21.15                        & 21.15               \\
3205.01    & 0            & 1                          & 57.98                        & 57.98               \\ \hline
Mean &	69.6 &	6.13 &	426.10	& 65.84 \\
Median & 13.5 & 5 & 264.94 & 65.20 \\
Std	& 187.72 & 5.84	& 452.14 & 27.11 \\ \hline
\end{tabular}
}
\caption{Data Statistics. The number of aggressive behavior onsets, the total observation time across all observation sessions, and the total number of observation sessions, per patient.}
\label{apptab:datastatistic}
\end{table}

\begin{table}[h]
\centering
\begin{tabular}{lccc}
\hline
\textbf{Process} & \textbf{History} & \textbf{Per event} & \textbf{Per session} \\
\hline
Homogeneous Poisson        & No              & $O(1)$ & $O(N)$  \\
NHPP (power law intensity) & No              & $O(1)$ & $O(N)$  \\
Hawkes (power law kernel)  & Yes             & $O(N)$ & $O(N^2)$ \\
Hawkes (exponential kernel)& Yes (recursive) & $O(1)$ & $O(N)$  \\
Hawkes ($K$ exp.\ kernels) & Yes (recursive) & $O(K)$ & $O(NK)$ \\
\hline
\end{tabular}
\caption{Time complexity of conditional intensity evaluation for temporal point processes with $N$ observed events per session.}
\label{apptab:intensity_complexity}
\end{table}

\begin{table}[t]
\centering
\small
\begin{tabular}{lcc}
\hline
\textbf{Model} & \textbf{Per session (s)} & \textbf{Per event (s)} \\
\hline
Hawkes (exponential kernel)    & $0.0165$ & $0.97 \times 10^{-3}$ \\
Hawkes (2 exponential kernels) & $0.0215$ & $1.26 \times 10^{-3}$ \\
Hawkes (power law kernel)      & $0.0246$ & $1.44 \times 10^{-3}$ \\
\hline
\end{tabular}
\caption{Empirical runtime of conditional intensity evaluation across Hawkes process variants.}
\label{apptab:timing_results}
\end{table}

\clearpage

\section{Supplementary Figures}

\clearpage

\section{Supplementary Equations}
\begin{align}
    \hat{K}(t) = \frac{T}{J^2} \sum_{j=1}^{J} \sum_{i\neq j} w_{ij} \mathbbm{1}(|t_i - t_j| \leq t) \label{appeq:ripleyk}
\end{align}
where $\frac{T}{J}$ is number of onsets per unit time, $J$ is the number of onsets observed in an observation session, $w_{ij}$ is an edge correction taking values $w_{ij}=1$ if $|t_i - t_j| \leq \min(t_i,T-t_i)$ and $w_{ij}=2$, otherwise.

\end{document}